\documentclass[ reprint,showpacs,
 superscriptaddress,
 pra,
]{revtex4-1}
\usepackage{amsfonts}
\usepackage{mathrsfs,amssymb}
\usepackage[top=1in, bottom=1in, left=1in, right=1in]{geometry}
\usepackage{amsmath}
\usepackage{epsfig}
\usepackage{bm}
\usepackage{graphicx}
\usepackage{bmpsize}
\usepackage{grffile}
\usepackage{cancel}
\usepackage{cases}
\usepackage{mhchem}
\usepackage{mathtools}
\usepackage{cool}

\begin{document}

\title{Exact solution of gradient echo memory and analytical treatment of gradient frequency comb}
\author{Xiwen Zhang}
\email{xiwen@physics.tamu.edu}
\affiliation{Department of Physics and Astronomy, Texas A\&M University, College Station, Texas 77843, USA}
\date{\today }

\begin{abstract}
Gradient echo memory (GEM) stores and retrieves photon wave packet in forward direction with high efficiency and fidelity using photon-echo mechanism. It is an important technique for quantum memory applications. By breaking the continuity of the gradient absorption structure, the scheme becomes gradient frequency comb (GFC), which is a hybrid of GEM and atomic frequency comb (AFC). To elucidate the non-trivial field-atom evolution of gradient echo, we derive its exact analytical solution in a medium of arbitrary optical thickness subjecting to any linear gradient of transition frequency, and discuss its physical processes. Based on this solution, we further suggest and analysis two types of GFC, one with stepwise gradient and the other with discontinuous gradient.
\end{abstract}


\maketitle
\section{Introduction}
Quantum information processing~\cite{DiVincenzo00}, suggested by Feynman and other pioneers~\citep{Feynman82, Deutsch85} in 1980s, has been one of the primary driving forces of a wide scope of quantum physics. One of its essential element is quantum memory~\cite{Lvovsky09, Simon10, Tittel10, Bussieres13} of a qubit, which lies in the heart of quantum communications~\cite{Briegel98} and quantum computations~\cite{Perez-Delgado11}, and provides a way to realize linear logic gates~\cite{Campbell14}, on-demand single-photon sources and single photon detectors~\cite{Imamoglu02,James02}, precision measurements~\cite{Honda08,Appel08,Appel09}, etc. The developed quantum memory techniques can also be transferred to other applications, such as ultrasound detection~\cite{McAuslan12}, etc.

Single photon is considered to be the ideal ``flying qubit" for quantum information processing because of its fast transmission speed. Quantum memory of a single photon of a time-bin state~\citep{Brendel99} can be achieved in a number of ways. For example, electromagnetically induced transparency (EIT)~\citep{Chaneliere05, Eisaman05, Novikova07, Choi08, Heinze13, Chen13} and off-resonant Raman~\cite{Reim11, Reim10, Reim12, Bustard13, England13} schemes use an optimized shaped-in-time control field to adiabatically map the single-photon wave packet into and out from long-lived atomic coherence between two ground states. A storage loop~\cite{Pittman02} stores a single photon in a loop with on-pseudodemand retrieval controlled by a high-speed switcher. Atomic frequency comb (AFC)~\cite{Afzelius09, deRiedmatten08, Clausen12, Zhou13, Chaneliere10, Saglamyurek11, Sabooni10, Akhmedzhanov13, Timoney12, Lauritzen11, Saglamyurek15} utilizes a spectrally periodic absorption structure, produced by spectrum tailoring on a wide inhomogeneous broadening, to store a single-photon wave packet in a medium based on photon-echo mechanism. Another scheme based on photon-echo mechanism is gradient echo memory (GEM)~\cite{Hetet08, Moiseev01, Liao14GEM}, which stores a photon by controlling the rephasing condition of the polarization via manipulating an artificial, continuous, distributed-in-space inhomogeneous broadening.

In GEM, an external dc electric or magnetic field with gradient along the longitudinal direction is applied to a medium with narrow transition linewidth (see Fig. \ref{FigGEM}). Due to Stark or Zeeman effect, the atomic transition frequency is shifted, acquiring a spatial dependence along the path of photon propagation. This is equivalent to an artificial inhomogeneous broadening. Therefore, different frequency components of the input photon are resonantly absorbed at different longitudinal locations of the medium, and the excited polarizations quickly get out of phase, resulting in a destructive interference to suppress the field's reemission into space. The on-demand retrieval is achieved by reversing the gradient of the external electric or magnetic field, which corresponds to an inverse of the inhomogeneous broadening. This will enforce a rewind of the phases of the polarizations back to their original statuses, upon which a photon echo emerges from the medium. GEM scheme can be implemented in both two-level and three-level~\cite{Hetet08EXP3level} systems, and has been realized in many materials, including rare-earth-doped crystals~\cite{Hetet08EXP, Hedges10, Lauritzen10}, warm~\cite{Hetet08EXP3level, Hosseini09, Hosseini11, Higginbottom12, Glorieux12} and cold~\cite{Sparkes13} atoms.

The evolution equation of GEM can be reduced from Maxwell-Bloch equation to the following form:
\begin{align}
\frac{\partial }{\partial z}a(z,t)& =g^{\ast }NS(z,t),  \label{GEMa} \\
\frac{\partial }{\partial t}S(z,t)& =-(\gamma -i\beta z)S(z,t)-ga(z,t), \label{GEMS}
\end{align}
where $a(z,t)$ is the slowly varying amplitude of the single-photon annihilation operator and $S(z,t)$ is the slowly varying part of the collective atomic coherence operator, $g$ is the field-atom coupling constant, $N$ is the atomic density, $\gamma$ is the decoherence rate and $\beta$ is the frequency gradient.

Although the mechanism of GEM scheme is conceptually simple, its field-atom dynamics described by Eqs. (\ref{GEMa}) and (\ref{GEMS}) is very nontrivial. The longitudinal distribution of the inhomogeneous broadening, i.e., the $z$ dependence of the resonant condition, is essential for GEM. For a uniformly distributed inhomogeneous broadening, after reversing the resonant condition, the forward retrieval efficiency is limited to $54\%$ by reabsorption process~\cite{Sangouard07}. But in GEM, since the resonant condition is assigned along the path of photon propagation, the polarization rephasing is accompanied by a phase matching process (see Sec. \ref{GEM_sub2}), so that the reabsorption of the field during forward retrieval does not strongly confine the efficiency of the echo, allowing nearly $100\%$ of the input energy to be recalled with high fidelity~\cite{Hetet08EXP}. However, such inhomogeneity along propagation leads to a complicated field and atomic evolution. Theoretically, some analytical treatments of GEM have been done, including generalized investigation of time-reversible atom-light interaction~\cite{Moiseev11}, analytical calculation for narrow-band input signal~\cite{Longdell08,Moiseev08}, analysis of GEM equation in spatial-Fourier space~\cite{Hetet08}, and GEM solution in terms of a response to the excited coherence~\cite{Hush13}. However, so far the analytical solution of gradient echo in real space and time in terms of an arbitrary input signal still remains unknown.

In this paper, we give the exact solution of GEM evolution equation (\ref{GEMa}) and (\ref{GEMS}) during storage and retrieval for an arbitrary linear gradient without approximations. The solution will be written in its final form, i.e., in terms of a response to the input signal. In such a way, the process of echo formation in a medium subjecting to the frequency gradient
  \begin{equation}
  \beta(t)=\left\{
  \begin{array}{@{}ll@{}}
    \beta, & \text{if}\ t\leqslant 0 \\
    \beta^\prime, & \text{if}\ t>0
  \end{array}\right.
\end{equation}
becomes clear. This solution describes not only GEM, but also quantum memory schemes based on phase matching control (PMC)~\citep{Kalachev11, Zhang13, Kalachev13, Zhang14LP, Clark12} and control field spatial chirp~\cite{Zhang14PRA}, which demonstrate the same memory performance as GEM without requiring the existence of Stark or Zeeman effect in the material. Moreover, the gradient absorption of a single photon is also related to the study of single-photon superradiance~\cite{Dicke54, Svid08a, Scully09, Rohlsberger10, Svid10, Svidzinsky15}. So the solution and mathematical treatment presented in this paper may provide insights, for example, into the problem of the preparation and control of the timed Dicke state~\cite{Scully06,Adams11}.

The storage process of GEM (i.e., a gradient absorption without reversing $\beta$) is akin to a recently proposed quantum memory scheme based on gradient frequency comb (GFC)~\cite{Zhang15TBPgamma, Zhang15TBP}, which is a combination of AFC and GEM schemes. AFC has been implemented in rare-earth-doped crystals for heralded single photon memory~\cite{Rielander14}, quantum entanglement~\cite{Usmani12}, time-bin~\cite{Gundogan15} and polarization~\cite{Zhou12, Gundogan12, Clausen12} qubit storage, and lots of other applications. It stores a photon via the periodically occurring constructive interference during the evolution of the polarizations in the medium due to the beating of the spectral comb teeth. This spectral comb can be created by, for example, a series of pulse pairs~\cite{deRiedmatten08} applied to a wide inhomogeneous broadening of the excited state. In the same way, GFC stores a photon based on also this mechanism as in AFC, except that its spectral comb teeth are distributed along the longitudinal direction similar to GEM scheme but in a discrete manner. So it can be viewed as a discrete GEM without reversing the frequency gradient. Then a natural question is, how does the type of discreteness of the frequency gradient affect the emitted echoes? Specifically, we consider two kinds of discreteness in GFC scheme: The first one is a ``stepwise" version [Fig. \ref{FigGFC} (a)], meaning that each comb tooth corresponds to a single-frequency absorption line with linewidth determined by the decoherence rate of the system. Here the word ``gradient" only makes sense when viewing all the comb teeth together [Fig. \ref{FigGFC} (b)]. The second type is a ``discontinuous" gradient frequency [Fig. \ref{FigGFC} (c)], meaning that each comb tooth itself has a gradient inside; and even if the decoherence rate drops to zero, every one has a finite spectral width determined by its internal gradient and physical thickness along the gradient [Fig. \ref{FigGFC} (d)]. In this paper, we analytically solve the two types of GFC, and compare their echoes in different regimes.

This paper is organized as follows: In Sec. \ref{GEM}, the exact analytical solution of the GEM evolution equation is derived. We first solve for the storage process and compare the result with that of a flat single-frequency absorption (zero gradient) in Sec. \ref{GEM_sub1}, and then solve for the retrieval process and discuss the result in Sec. \ref{GEM_sub2}. In Sec. \ref{GFC}, we introduce the model of stepwise GFC and discontinuous GFC, and give their analytical analysis for comparison. We first derive the approximate solution for the GFC echoes in an optically thin medium based on the result in Sec. \ref{GEM_sub1}, and then derive the expression for the first several GFC echoes for arbitrary optical thickness and discuss the first echo optimization condition in Sec. \ref{GEM_sub2}. In Sec. \ref{Conclusion}, we discuss the result and conclude the article.

\section{The exact solution of gradient echo}
\label{GEM}
\begin{figure}[h]
\begin{center}
\epsfig{figure=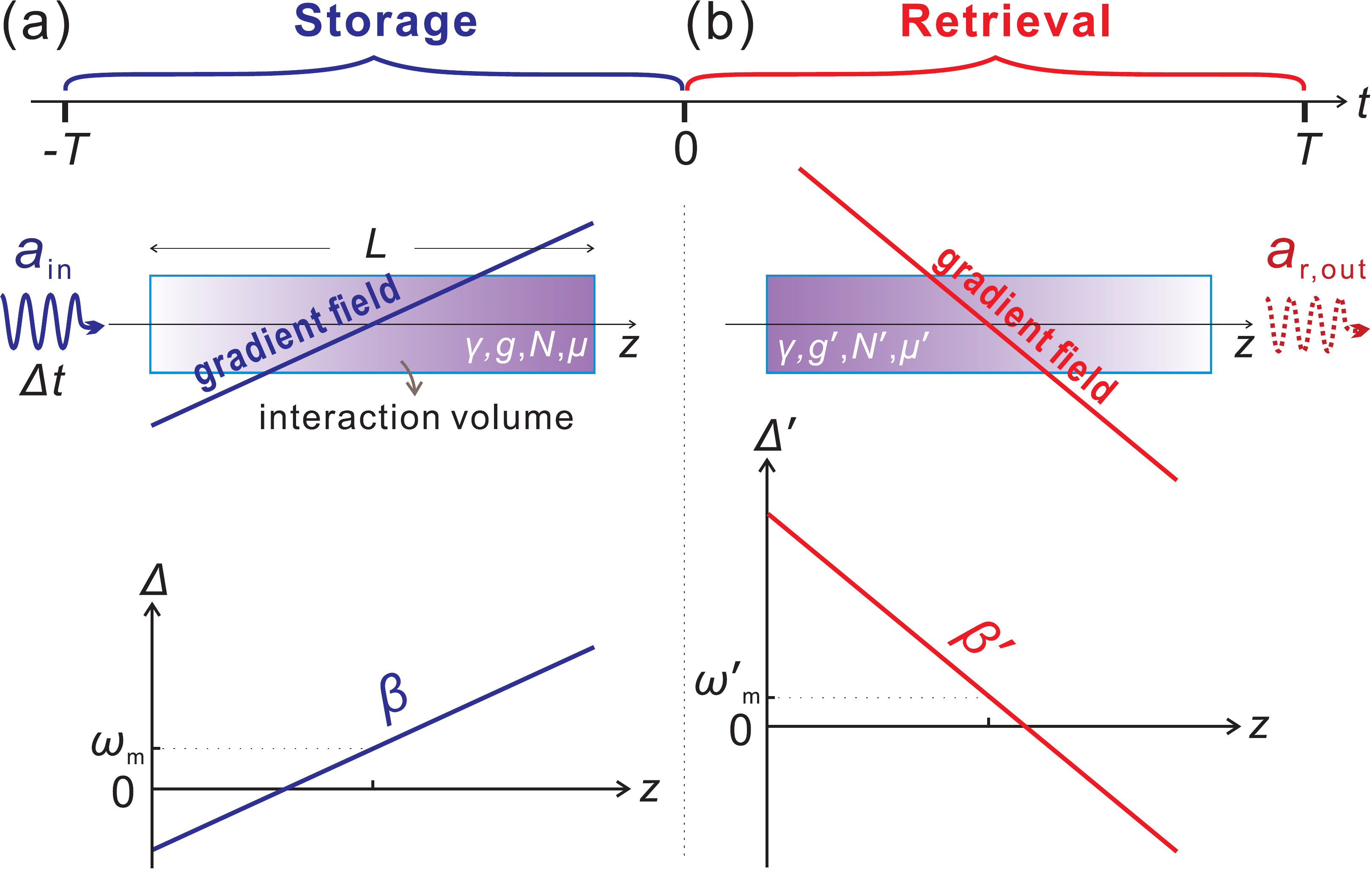, width=7.87cm}
\end{center}
\caption{(Color online) Illustration of GEM scheme. (a) Storage process ($t\in[-T,0]$). (b) Retrieval process ($t\in[0,T]$). }\label{FigGEM}
\end{figure}

In a GEM experiment, a single-photon (or weak signal field) wave packet with slowly varying amplitude $a_\text{in}(t)$ and full-width-half-maximum (FWHM) field duration $\Delta t$ propagating along the longitudinal direction $\hat{z}$ enters a resonant two-level (or three-level) medium of length $L$. All the atoms in the medium are identical, initially staying in their ground states and mainly remaining unexcited during the whole storage and retrieval processes. An external gradient electric or magnetic field is applied to the medium to create a one-photon (or two-photon) detuning varying linearly as a function of position: $\Delta = \beta z$, where $\beta $ is the frequency gradient along the photon propagation direction [see Fig. \ref{FigGEM} (a)]. During retrieval, the gradient switches to a different value $\beta \rightarrow \beta ^{\prime }$, as shown in Fig. \ref{FigGEM} (b). Very often $\beta ^{\prime }=-\beta $ is chosen to symmetrically reverse the photon detuning. Denoting the slowly varying amplitude of the single-photon annihilation operator as $a(z,t)$ and slowly varying amplitude of the collective atomic coherence operator as $S(z,t)$, the evolution equation for the field and atom within $z\in \lbrack -L/2,L/2]$ in the long-pulse regime ($c \Delta t \gg L$, $c$ is the speed of light in vacuum) is given by Eqs. (\ref{GEMa}) and (\ref{GEMS}).



In this section, the storage and retrieval processes are in time intervals $t\in [-T,0]$ and $t\in [0,T]$ respectively (Fig. \ref{FigGEM}), where $T$ is the duration of the storage or retrieval time window. The relevant parameters are defined as the following: $\gamma$ is the single-atom decoherence rate of the optical (spin) transition in the two-level (three-level) system, $L$ is the medium length, $\beta$ is the frequency gradient, $g$ is the field-atom coupling constant, $N$ is the atomic density, $\zeta = 2|g|^2 NL / \gamma$ is the optical thickness of the medium, $\mu = |g|^2N/\beta$, $\omega_m$ is an additional frequency shift on top of the frequency gradient. The parameters $\beta^\prime$, $g^\prime$, $\mu^\prime$ and $\omega^\prime$ carrying ``$^\prime$" indicate that they are quantities specifically for retrieval. The subscript ``s'' (``r") of a function denotes the storage (retrieval) process.

The major figures of merit for a quantum memory scheme are efficiency $\eta $ and fidelity $\mathscr{F}$~\cite{Zhang14LP}, which we define as follows:
\begin{equation}
\eta =\frac{N_{\text{out}}}{N_{\text{in}}},
\end{equation}
where $N_{\text{in}}=\int_{-T }^{0}dt\,\langle a_{\text{in}}^{\dag}(t)a_{\text{in}}(t)\rangle $ and $N_{\text{out}}=\int_{0}^{T }dt \langle a_{\text{out}}^{\dag }(t)a_{\text{out}}(t)\rangle $. Here $a_{\text{in}}(t)$ and $a_{\text{out}}(t)$ are the slowly varying part of the input and output field annihilation operators respectively. The fidelity can be defined as
\begin{equation}
\mathscr{F}=\frac{1}{N_{\text{in}}N_{\text{out}}}\left\vert
\int_{0}^{T }dt\,\langle a_{\text{in}}^{\dagger }(\bar{t}-t)a_{\text{out
}}(t)\rangle \right\vert ^{2},  \label{FidelityF}
\end{equation}
where $\bar{t}-t$ takes into account the time reversal and possible delay of the output field in optical dense medium (see Sec. \ref{GEM_sub2}). For comparison, we also introduce amplitude preservation $\mathscr{A}$ to describe the conservation of the temporal shape of the echo:
\begin{equation}
\mathscr{A}=\frac{1}{N_{\text{in}}N_{\text{out}}}\Bigg (\int_{0}^{T
}dt\,\left\vert \langle a_{\text{in}}^{\dagger }(\bar{t}-t)a_{\text{out}
}(t)\rangle \right\vert \Bigg )^{2}.  \label{FidelityA}
\end{equation}

A more general form of the evolution equation (\ref{GEMa}) and (\ref{GEMS}) is:
\begin{align}
\frac{\partial }{\partial z}a(z,t)& =g^{\ast }Ns(z,t)e^{i\phi (z,t)},
\label{EQa1} \\
\frac{\partial }{\partial t}s(z,t)& =-\gamma s(z,t)-ga(z,t)e^{-i\phi (z,t)},
\label{EQs1}
\end{align}
where
\begin{align}
\phi (z,t) &=\int^{t}d\tau \Delta (z,\tau )+\phi _{0}(z),  \label{phi1} \\
s(z,t) &=S(z,t)e^{-i\phi (z,t)},
\end{align}
$z\in[-L/2,L/2]$. Here $\phi$ or $\Delta$ can be various functions of $z$ and $t$. For example, a stepwise-in-space $\phi$ (along with corresponding discrete interaction volume) reduces the system to GFC scheme~\cite{Zhang15TBPgamma}. In this section we will restrict ourselves to a linear phase in space and time:
\begin{equation}
\phi (z,t)=\beta zt+\omega _{m}t,
\end{equation}
i.e., $\Delta =\beta z + \omega_m$. Here $\omega_m$ is a frequency shift in addition to the space dependent detuning $\beta z$, which will shift the echo central frequency as discussed in Ref.~\cite{Buchler11} and experimentally demonstrated in Refs.~\cite{Hosseini11, Sparkes12}. Another reason for including this frequency shift $\omega_m$ is that it corresponds to the central frequency of a comb tooth in the GFC scheme (see Sec. \ref{GFC_sub1}).

The GEM evolution equation in the form of Eqs. (\ref{EQa1}) and (\ref{EQs1}) reflects the mechanism of quantum memory schemes based on PMC~\cite{Zhang14LP, Zhang13, Kalachev13} and control field spatial chirp~\cite{Zhang14PRA}. For example, in PMC, the input single photon is mapped into and out from the spin wave excitation according to externally controlled phase matching condition, which is described exactly by this phase factor $\phi(z,t)$. The form of Eqs. (\ref{EQa1}) and (\ref{EQs1}) is also more convenient for deriving the exact analytical solution of gradient echo. In order to do this, let us change the spatial domain into $z\in[0,L]$:
\begin{align}
\frac{\partial }{\partial z}a(z,t)& =g^{\ast }Ns(z,t)e^{i\beta z t}e^{-i(\beta L/2-\omega _{m})t},  \label{EQa2} \\
\frac{\partial }{\partial t}s(z,t)& =-\gamma s(z,t)-ga(z,t)e^{-i\beta z t}e^{i(\beta L/2-\omega _{m})t},  \label{EQs2}
\end{align}
and then apply Laplace transformation method over $z$. The solution of Eqs. (\ref{EQa2}) and (\ref{EQs2}) in Laplace domain is given by Eqs. (\ref{App_ap}) and (\ref{App_sp}) in Appendix \ref{AppSec_GeneralSln}.

\subsection{Storage}
\label{GEM_sub1}
Before storage, there is no collective atomic coherence in the medium, so the initial condition is $s_{s}(z,-T)=0$. The boundary condition is given by the incoming signal field: $a_{s}(z=0,t)=a_{\text{in}}(t)$. Subjecting to these initial and boundary conditions, the exact solution of Eqs. (\ref{EQa2}) and (\ref{EQs2}) during the storage process is obtained in Appendix \ref{AppSec_Storage} using inverse Laplace transformation of Eqs. (\ref{App_ap}) and (\ref{App_sp}). The results are given by Eqs. (\ref{App_as1}) and (\ref{App_ss1}).
Since the gradient absorption can be considered as the light-matter interaction due to a distributed single-frequency absorption, it is illustrative to compare the field-atom evolution of these two systems. The latter corresponds to the case when $\beta=0$, whose solution is given by Eqs. (\ref{App_as1_beta=0}) and (\ref{App_ss1_beta=0}). The results can be summarized as follows:
\begin{align}
a_{s}(z,t\leqslant 0)&=\int_{-T}^{0}d\tau a_{\text{in}}(\tau )f_{s}(z,t-\tau ), \label{as2} \\
s_{s}(z,t\leqslant 0)&=e^{-i\beta \left(z-\frac{L}{2}\right)t} \times \notag \\
&\int_{-T}^{0}d\tau a_{\text{in}}(\tau )e^{-i\omega_m \tau}h_{s}(z,t-\tau ), \label{ss2}
\end{align}
where
\begin{align}
& f_{s}^{(\beta \neq 0)}(z,t) =\delta (t)-\mu \beta ze^{-i\left( \beta \frac{L}{2}-\omega_{m}\right) t}e^{-\gamma t} \times \notag \\
& \qquad \qquad \qquad \qquad \quad \Hypergeometric{1}{1}{i\mu +1}{2}{i\beta z t }\Theta (t),  \label{GEMstorageResp} \\
& f_{s}^{(\beta =0)}(z,t)=\delta (t)-|g|^{2}Nze^{i\omega _{m}t}e^{-\gamma t} \times \notag \\
& \qquad \qquad \qquad \qquad \quad \tilde{J}_{1}(|g|^{2}Nzt)\Theta (t),\label{NoGRADstorageResp}
\end{align}
\begin{align}
& h_{s}^{(\beta \neq 0)}(z,t) =-ge^{-i\beta \frac{L}{2}t}e^{-\gamma t}\,\Hypergeometric{1}{1}{i\mu +1}{1}{i\beta z t}\Theta (t), \label{GEMstorageCoherenceResp} \\
& h_{s}^{(\beta =0)}(z,t) =-ge^{-\gamma t}J_{0}(2\sqrt{|g|^{2}Nzt})\Theta (t), \label{NoGRADstorageCoherenceResp}
\end{align}
and $z\in[0,L]$. Here $_{1}F_{1}$ is the Kummer confluent hypergeometric function, $\tilde{J}_{1}(x) = J_{1}(2\sqrt{x})/\sqrt{x}$, where $J_\nu$ is the $\nu ^\text{th}$ order Bessel function, and $\Theta (t)$ is the Heaviside step function.
$f_{s}(z,t)$ is the response function of the GEM system during storage. Its second term tells how a weak input field gets absorbed by a medium with longitudinally distributed, linear, gradient transition frequency.


Recalling that $\mu \beta = |g|^2 N$, it can be seen from Eqs. (\ref{GEMstorageResp}) and (\ref{NoGRADstorageResp}) that the difference between a gradient absorption and a flat single-frequency absorption lies in the functions $e^{-i\beta tL/2}\,\Hypergeometric{1}{1}{i\mu+1}{2}{i\beta z t }$ and $\tilde{J}_{1}(|g|^{2}Nzt)$, which are plotted Fig. \ref{FigResponseFunc} in Appendix \ref{AppSec_Storage}. This difference manifests itself in the expansion these two functions:
\begin{align}
& \Hypergeometric{1}{1}{i\mu + 1}{2}{i\beta z t} e^{-i\beta \frac{L}{2} t} = \notag \\
& \quad e^{-i\beta \frac{L}{2} t} \sum_{n=0}^{\infty }\frac{\left( 1-i\frac{1}{\mu }\right) \cdots \left( 1-i\frac{n}{\mu }\right) }{(n+1)!}\frac{\left(-|g|^{2}Nzt\right) ^{n}}{n!},  \label{1F1expand} \\
& \tilde{J}_{1}(|g|^{2}Nzt)  =\sum_{n=0}^{\infty }\frac{1}{(n+1)!}\frac{(-|g|^{2}Nzt)^{n}}{n!}.  \label{J1expand}
\end{align}
Comparing Eqs. (\ref{1F1expand}) and (\ref{J1expand}), it can be seen that in the limit $\beta \rightarrow 0$, i.e., $\mu \rightarrow \infty $, the response function (\ref{GEMstorageResp}) reduces to (\ref{NoGRADstorageResp}) as expected.
From these two expansions, to the first order smallness of $\beta $ we have:
\begin{align}
& \Hypergeometric{1}{1}{i\mu + 1}{2}{i\beta z t} e^{-i\beta \frac{L}{2} t} \notag \\
& \qquad \qquad \approx  \left( 1+i\beta\frac{ z-L}{2}t\right) \tilde{J}_{1}(|g|^{2}Nzt). \label{F11Expend2}
\end{align}
In fact, using the integral representation of $_1F_1$, $e^{-i\beta t L/2 } \Hypergeometric{1}{1}{i\mu + 1}{2}{i\beta z t}$ can be expanded as:
\begin{align}
& \Hypergeometric{1}{1}{i\mu + 1}{2}{i\beta z t } e^{-i\beta \frac{L}{2} t} \notag \\
= & \frac{e^{i\beta \left(z - \frac{L}{2} \right) t}\mu ^{1-i\mu }}{\Gamma (1-i\mu )}\int_{0}^{\infty}d\xi e^{-\mu (\xi +i\ln \xi )} \tilde{J}_{1}(i|g|^{2}Nzt\xi ), \label{1F1integral2}
\end{align}
which tells how a gradient absorption built up from a flat single-frequency absorption response.

\begin{figure}[h]
\begin{center}
\epsfig{figure=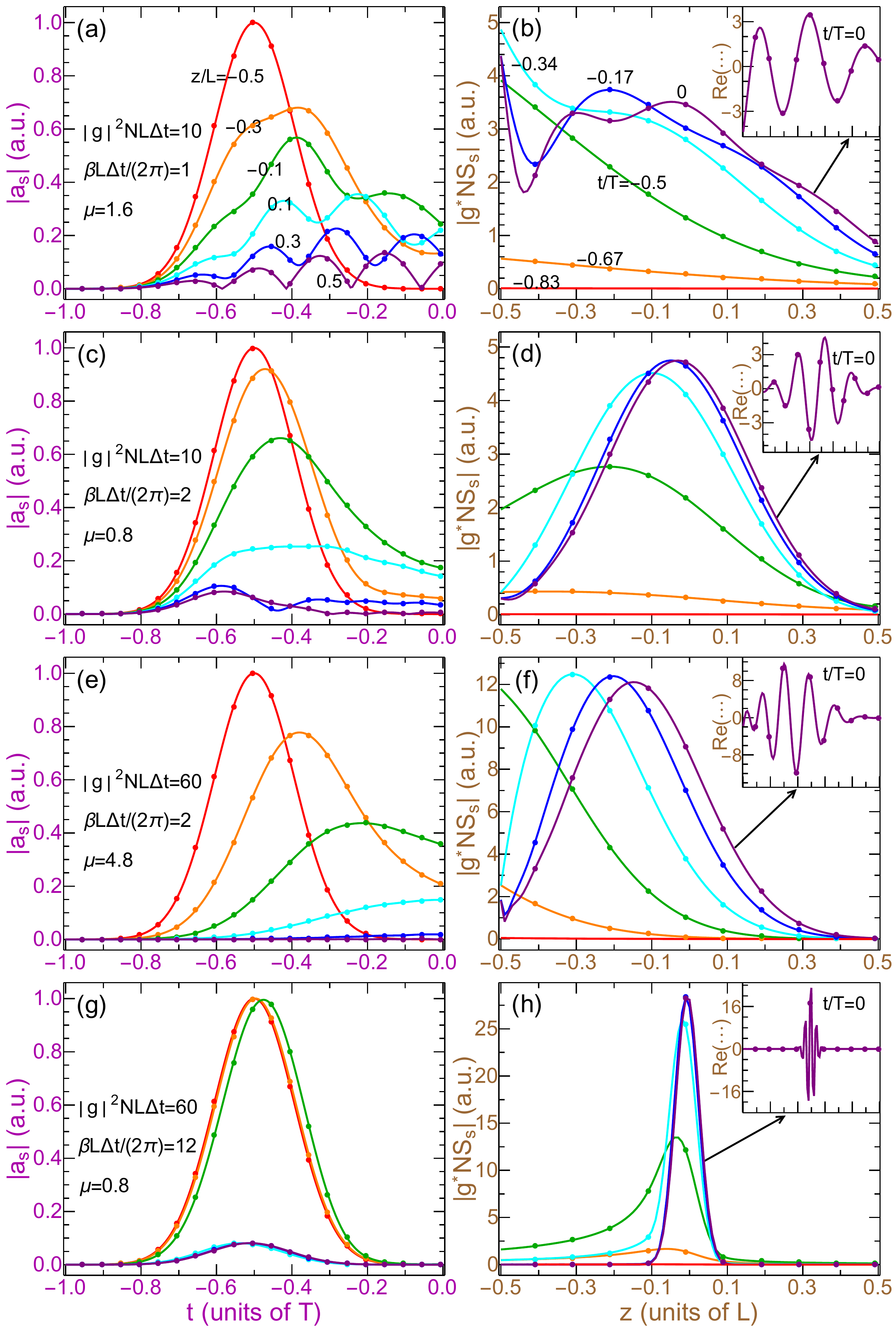, width=7.95cm}
\end{center}
\caption{(Color online) Analytical solution (dotted) based on Eqs. (\ref{as2}), (\ref{ss2}) and numerical simulation (solid) of Eqs. (\ref{GEMa}), (\ref{GEMS}) for GEM storage process. The input single photon has a Gaussian waveform (red curves in the left panels) of duration $\Delta t/T = 0.25$ peaked at $t_\text{in}/T = -0.5$. Spatial variable $z$ is converted to the interval $[-L/2, L/2]$. (a, c, e, g) Absolute value of the field $|a_s(z,t)|$ as a function of time at different locations of the medium. The colors of red, orange, dark green, cyan, blue and purple represent $z/L=-0.5,-0.3,-0.1,0.1,0.3,0.5$ respectively. (b, d, f, h) Absolute value of the collective coherence $|g^* N s_s(z,t) e^{i\beta z t}|$ as a function of space at different times. The colors of red, orange, dark green, cyan, blue and purple represent $t/T=-5/6,-2/3,-1/2,-1/3,-1/6,0$ respectively. The inset show its real value at the end of the storage, i.e., t/T = 0. The figures are obtained under parameters $\gamma = 0$ and $\omega_m=0$. The value of $\mu$ is $1.6$ in (a, b), $0.8$ in (c, d), (g, h), and $4.8$ in (e, f). }\label{FigStore}
\end{figure}

In Fig. \ref{FigStore} we plot the gradient absorption process of an input photon (with Gaussian waveform) of duration $\Delta t$ in a medium of length $L$ when $\omega_m=0$. It shows that the analytical solution (\ref{as2} - \ref{GEMstorageResp}), (\ref{GEMstorageCoherenceResp}) and numerical simulation of the evolution equation (\ref{GEMa}) and (\ref{GEMS}) [or (\ref{EQa2}) and (\ref{EQs2})] agree with each other. From Fig. \ref{FigStore} (a-d, g-h), for fixed $|g|^2N$ the gradient absorption experiences more amplitude and phase modulation in the case of small storage bandwidth $\beta L$ due to the cutoff of the absorbed frequency components [Fig. \ref{FigStore} (a, b)], and for fixed $\mu$ a large storage bandwidth $\beta L$ compresses the collective coherence distribution into a narrow region corresponding to the spectrum of the input field [Fig. \ref{FigStore} (h)]. So $\beta$ determines the spatial width of the polarization distribution in the medium.

Comparing Fig. \ref{FigStore} (c, e), since $2|g|^2NL/\gamma$ is the optical thickness, for enough storage bandwidth a larger $|g|^2N$ indicates a faster absorption of the input field.

Comparing Fig. \ref{FigStore} (d, f, h), we see that for sufficient absorption bandwidth and optical density, $\mu$ determines the central position of the polarization distribution in the medium. Indeed, if we divide the medium into a number of spectrally resolved units according to $\beta L / (2\gamma)$, $\mu$ is proportional to the optical thickness of each of such units. In an optically dense medium, the absorption width is broadened by optical thickness compared with the single-atom absorption linewidth, so the input field gets absorbed near the entrance (rather than in the center) of the medium even though the resonant absorption frequency in this region does not exactly match the input spectrum [Fig. \ref{FigStore} (f)]. But when $\mu$ gets smaller, this central position of the polarization distribution is pushed back to the center of the medium because the optical density of each spectrally resolved unit is ``diluted" by large storage bandwidth [Fig. \ref{FigStore} (h)].

Next, let us look at two special cases of the gradient absorption. In the first example, we consider an infinite broad-band input field:
\begin{align}
&a_{\text{in}}(t)=\delta (t-t_{\text{in}}), \text{ } -T < t_{\text{in}} < 0. \label{DeltaInput}
\end{align}
Then from Eqs. (\ref{as2}) and (\ref{ss2}), we have the output field (field not being absorbed by the medium) and the collective coherence after storage as
\begin{align}
a_\text{s,out}& (t)=\delta (t-t_{\text{in}})-\mu \beta Le^{-i \left( \beta \frac{%
L}{2} -\omega_m \right)(t-t_{\text{in}})} \times \notag \\
& e^{-\gamma (t-t_{\text{in}})}\,\Hypergeometric{1}{1}{i\mu +1}{2}{i\beta L ( t -t_\text{in} ) }\Theta (t-t_{%
\text{in}}),
\end{align}
\begin{align}
s_{s}(z,t =0)=-g&e^{i\left( \beta \frac{L}{2} - \omega_m \right) t_{\text{in}}}e^{\gamma t_{\text{in}%
}} \times \notag \\
& \qquad  \Hypergeometric{1}{1}{i\mu +1}{1}{-i\beta  t_\text{in} z},
\end{align}
in which $t \in [-T,0]$, $z\in [0,L]$.

\begin{figure}[h]
\begin{center}
\epsfig{figure=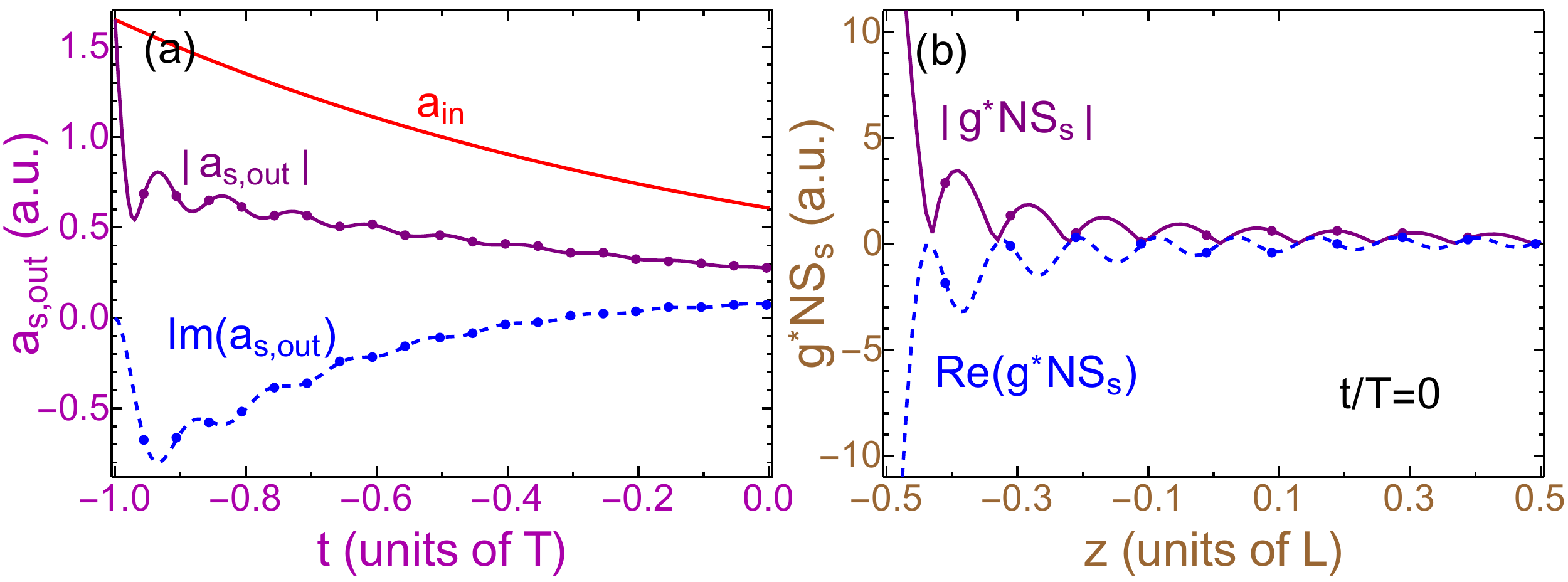, width=8.0cm}
\end{center}
\caption{(Color online) Analytical solution [dotted, based on Eqs. (\ref{aStorage_ExpDecay}) and (\ref{sStorage_ExpDecay})] and numerical simulation (solid and dashed lines) for GEM storage process with a quasi-monochromatic input single photon wave packet $a_{\text{in}}(t)=e^{-\gamma (t-t_{\text{in}})}$, $t_{\text{in}} = -T/2$. (a) The input field $a_\text{in}(t)$, and absolute value and imaginary part of the output $a_\text{s,out}(t)$ field as functions of time. (b) The absolute value and real part of the collective coherence $g^* N S_\text{s}(z,t=0)$ distributed in the medium after storage. The spatial variable is converted to $z \in [-L/2,L/2]$. The figures are obtained under parameters $\gamma = 1/T$, $|g|^2 N L T = 40$, $\beta L T = 50.3$, and $\mu=0.8$. }\label{FigStoreExpDecay}
\end{figure}

In the second example, we consider a narrow-band input field absorbed by a gradient system with an additional frequency shift $\omega_m$ on top of the resonant absorption bandwidth:
\begin{align}
&a_{\text{in}}(t)=e^{-\gamma (t-t_{\text{in}})}, \text{ } -T < t_{\text{in}} < 0, \label{ExpDecayInput} \\
&\omega_m=\beta L/2, \label{ExpDecayInputFreqShift}
\end{align}
with $t \in [-T,0]$. In such a case, the exact expressions of the output field and collective coherence after storage are
\begin{align}
a_\text{s,out}(t)=e^{-\gamma (t-t_{\text{in}})}\,\Hypergeometric{1}{1}{i\mu }{1}{i\beta L ( t +T )}, \label{aStorage_ExpDecay}
\end{align}
\begin{align}
s_{s}(z,t& =0)= - g T e^{\gamma t_{\text{in}}}\,\Hypergeometric{1}{1}{i\mu
+1}{2}{i\beta  T z }, \label{sStorage_ExpDecay}
\end{align}
respectively, where $z\in [0,L]$. This solution is plotted in Fig. \ref{FigStoreExpDecay}.

\subsection{Retrieval}
\label{GEM_sub2}
The retrieval time window is $t\in [0,T]$. Let us assume that right after storage, at $t=0$, the parameters of the system are switched in the following way: the frequency gradient $\beta \rightarrow \beta^{\prime }$, coupling constant $g \rightarrow g^{\prime }$, atomic density $N \rightarrow N^{\prime }$, $\mu \rightarrow \mu ^{\prime }=\left\vert g^{\prime}\right\vert ^{2}N^{\prime }/\beta ^{\prime }$, and the additional frequency shift $\omega_m \rightarrow \omega _{m}^{\prime }$. The initial condition is given by the collective coherence distribution determined by the result of the storage process. In usual GEM, this is $s_{r}(z,0)=s_{s}(z,0)$ where the subscript ``r" denotes the retrieval process. Since there is no input field during retrieval, the boundary condition is $a_{r}(z=0,t)=0$.

\begin{figure}[h]
\begin{center}
\epsfig{figure=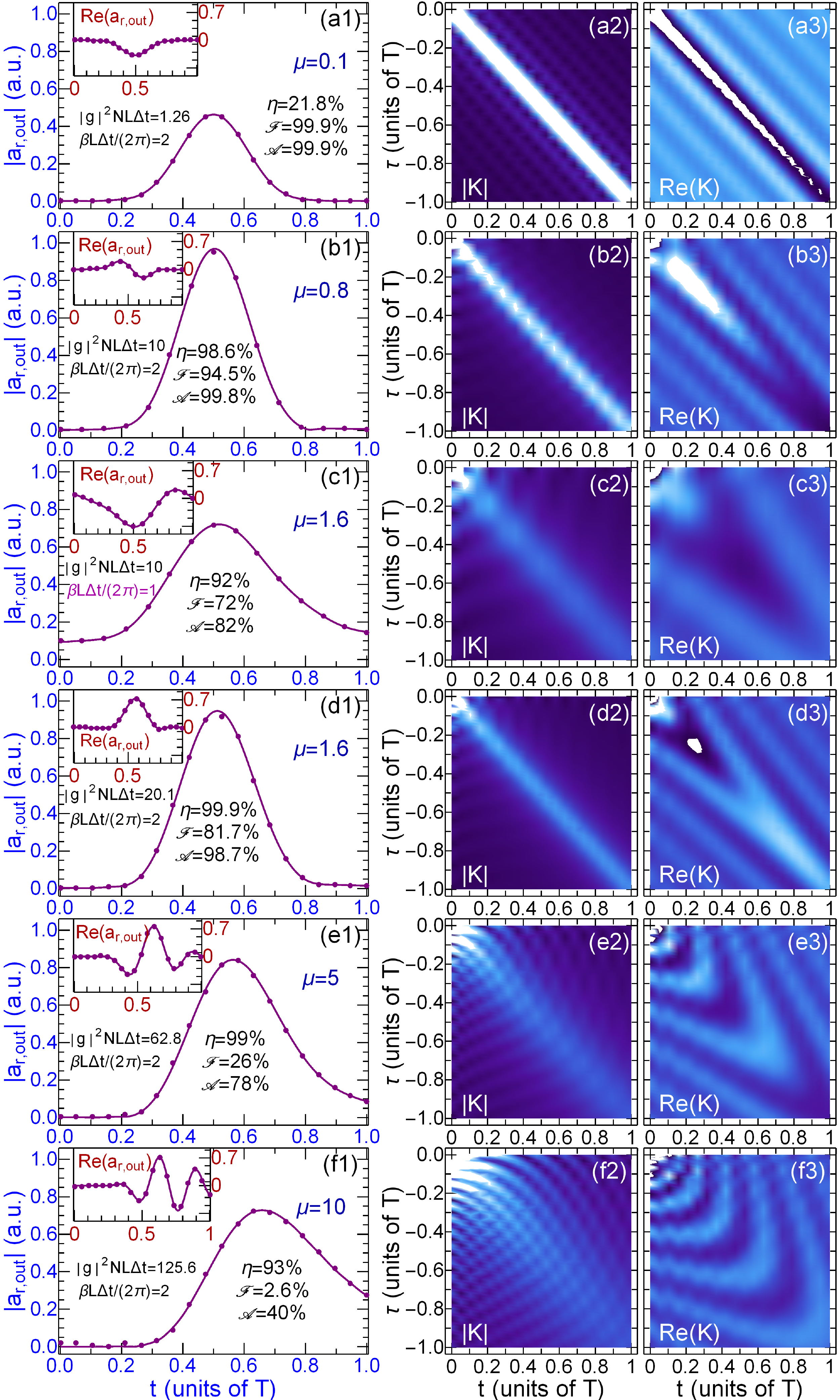, width=7.95cm}
\end{center}
\caption{(Color online) Analytical solution (dotted) based on Eqs. (\ref{ar3}), (\ref{Kkernel}) and numerical simulation (solid) of Eqs. (\ref{GEMa}), (\ref{GEMS}) for the output gradient echo $a_\text{r,out}(t)$ (left panels), as well as the corresponding absolute value (middle panels) and real part (right panels) of the integral kernel $K$, for different parameters shown in the figure. The incoming photon has a Gaussian temporal shape peaked at $t_\text{in}/T = -0.5$ with $\Delta t/T=0.25$. }\label{FigRetrieve}
\end{figure}

We first consider a typical GEM experiment, where only the frequency gradient is switched to the opposite during retrieval, meaning that $g^{\prime }=g$, $N^{\prime }=N$, $\beta ^{\prime}=-\beta $, $\mu ^{\prime }=-\mu $. Its exact analytical solution is given by Eqs. (\ref{App_ar2}) and (\ref{App_sr2}) in Appendix \ref{AppSec_Retrieval}.
From Eq. (\ref{App_ar2}), by setting $z=L$ in the retrieval field $a_r(z,t)$, the output GEM echo can be written into
\begin{align}
a_\text{r,out}(t\geqslant 0)&=\int_{-T}^{0}d\tau a_\text{in}(\tau) K(t,\tau), \label{ar3}
\end{align}
\begin{align}
K(& t,\tau) =-\mu \beta L e^{i\beta \frac{L}{2}\left( t+\tau \right) }e^{i(\omega _{m}^{\prime}t-\omega _{m}\tau )}e^{-\gamma (t-\tau )} \times \notag \\
& \Phi _{2}\left( i\mu +1,-2i\mu;2;-i\beta L\left( t+\tau \right) ,-i\beta L t\right) , \label{Kkernel}
\end{align}
where $t \in [0,T]$. Here the essential part of the integral kernel $K$ is $\Phi _{2}$, which is the Humbert double hypergeometric series. Some of this special function's properties are discussed in Appendix \ref{AppSec_Retrieval_sub1}. According to Eq. (\ref{App_Phi2_4}), it is straightforward to write the GEM output echo into a series expansion in terms of Gauss hypergeometric function $_{2}F_{1}$, as shown in Eq. (\ref{App_ar3}).
In Fig. \ref{FigRetrieve}, we plot the analytical solution and numerical simulation of the output gradient echoes (along with their integral kernels $K$) for different parameters and show that the two agree with each other.

As seen from the middle panels of Fig. \ref{FigRetrieve}, the modulus of the integral kernel $|K|$ behaves like a dirac delta function that peaks at $\tau = -t$ to select the corresponding moments from the input photon wave packet $a_\text{in}(\tau)$ in the integral (\ref{ar3}), which recovers the signal field in a time-reversed order. This is governed by the argument $i\beta z(t+\tau)$ in $\Phi_2$, as seen from Eq. (\ref{App_ar2}) [or (\ref{Kkernel}) for $z=L$], which is the total phase carried by a specific spatial mode of the polarization as a function of time $t$ during retrieval. From the time when the gradient is reversed ($t=0$), this phase starts to regress back towards its original value zero. The complete rephasing occurs at $t=-\tau$, which determines the moment of a constructive interference and manifests itself as the maximized point of $|K|$.

However, just the rephasing is not enough for understanding GEM, and the longitudinal dependence of the resonant condition plays a crucial role in the field evolution. Without this $z$-dependence, for a system with longitudinal-uniformly distributed inhomogeneous broadening, the forward echo after the reverse of this broadening contains at most $54\%$ of the input energy because of the reabsorption on its path during propagation~\cite{Sangouard07, Hetet08EXP}. But in GEM, since the argument $i\beta (t+\tau)z$ in Eq. (\ref{App_ar2}) is valid for all space in the medium, we should meanwhile treat $k_s=\beta \tau$ and $k_r = \beta^\prime t = -\beta t$ as the wave vectors characterizing the spatial mode of the polarization in the medium during storage and retrieval, respectively. This spatial mode of the polarization is caused simply by its free evolution of the phase due to position-dependent transition frequency. The gradient echo is mapped out from the collective excitation in the medium in a way that the retrieval procedure finds its correct spatial mode of the polarization in time according to the phase matching condition $k_r = k_s$. In this manner the echo avoids too much reabsorption during its evolution and demonstrates high efficiency larger than $54\%$.

Comparing Fig. \ref{FigRetrieve} (c) and (d), for fixed $\mu$, a smaller bandwidth $\beta L$ corresponds to a wider peak width of the kernel $|K|$. This is expected from Eq. (\ref{App_ar2}) [or (\ref{Kkernel})], since $\beta z$ enters $\Phi_2$ as a scaling factor in front of time variable.

Comparing Fig. \ref{FigRetrieve} (a, b, d-f), for fixed bandwidth $\beta L$ that covers the input spectrum, e.g. here $\beta L = 4\pi / \Delta t$, if $\mu \ll 1$, the optical thickness is too small to retain the input energy and therefore the storage efficiency remains very low, as seen from Fig. \ref{FigRetrieve} (a1). On the other hand, if $\mu \gg 1$, the linewidth of each transition frequency is broadened by large optical thickness, which smears out the phase matching condition (discussed in Sec. \ref{GEM}) and makes the peak of the integral kernel $|K|$ wide, as seen from Fig. \ref{FigRetrieve} (f2). It is then straightforward to see from Eq. (\ref{ar3}) that this will eventually delay the echo and deform its waveform.
This waveform deformation is quantified by the amplitude preservation $\mathscr{A}$. As shown by Fig. \ref{FigRetrieve} (f1), for large $\mu$, the gradient echo of an input Gaussian wave packet peaked at $t_\text{in}=-0.5T$ becomes right-shifted in time and asymmetric in shape, which reduces $\mathscr{A}$ down to $40\%$. Moreover, from the right panels of Fig. \ref{FigRetrieve} (a, b, d-f), it is seen that a large optical thickness also introduces a phase modulation. So the strong light-matter interaction not only drags and distorts the echo waveform, but also intensely modulates its phase. This phase modulation on the echo, shown in the insets of the left panels of Fig. \ref{FigRetrieve}, will rapidly decrease the memory fidelity $\mathscr{F}$. In order to keep high fidelity, $\mu$ should be kept small enough such that the phase variation of $K$ on the scale of input photon duration $\Delta t$ is insignificant, like in Fig. \ref{FigRetrieve} (a) and (b).

From the above analysis, in order to optimize the gradient echo, i.e., to maximize both efficiency and fidelity, the bandwidth $\beta L$ should cover the input bandwidth, and the parameter $\mu$ should be on the order of $1$:
\begin{align}
& \beta L \gtrsim 2\pi / \Delta t, \label{GEMOptCond1} \\
& \mu \sim 1. \label{GEMOptCond2}
\end{align}
Too small $\mu$ results in small efficiency, and too large $\mu$ results in small fidelity (as well as efficiency and amplitude preservation $\mathscr{A}$). An example of high efficiency and fidelity is shown in Fig. \ref{FigRetrieve} (b), with $\eta = 99\%$ and $\mathscr{F}=95\%$.

In a more general case, $g^\prime$ and $g$ can be different, and $\beta^\prime$ is not necessarily equal to $-\beta$ either. In Appendix \ref{AppSec_Retrieval}, we calculate the time and space evolution of the retrieval field and collective coherence for arbitrary $g^\prime$, $N^\prime$, $\beta^\prime$, $\mu^\prime$ and $\omega^\prime$. The result is given by Eqs. (\ref{App_ar1}) and (\ref{App_sr1}). GEM with additional frequency shift $\omega_m$, and different amounts of the frequency gradient between storage and retrieval, has been considered numerically in Refs.~\cite{Moiseev10,Buchler11}, and demonstrated experimentally in Refs.~\cite{Hosseini09, Hosseini11, Sparkes12}. The former (nonvanishing $\omega_m$ and/or $\omega_m^\prime$) leads to the echo frequency modulation, which is clearly seen in Eq. (\ref{App_ar1}). The latter causes the retrieval pulse stretching or compressing, due to the change of the speed of rephasing process, or the speed of reading out the spatial mode of the polarization. This can be seen by the phase matching condition $\beta^\prime t = \beta \tau$, which results in $t = (\beta/\beta^\prime) \tau$, meaning that $\Delta t_\text{out} = |\beta/\beta^\prime| \Delta t_\text{in}$. Meanwhile, since $\tau \leqslant 0 \leqslant t$, this phase matching condition is possible only if $\text{sign}(\beta^\prime) = -\text{sign}(\beta)$, unless additional amount of phase of the polarization is introduced by other means~\cite{Hosseini09, Zhang14PRA, Zhang15TBP}.

\begin{figure}[h]
\begin{center}
\epsfig{figure=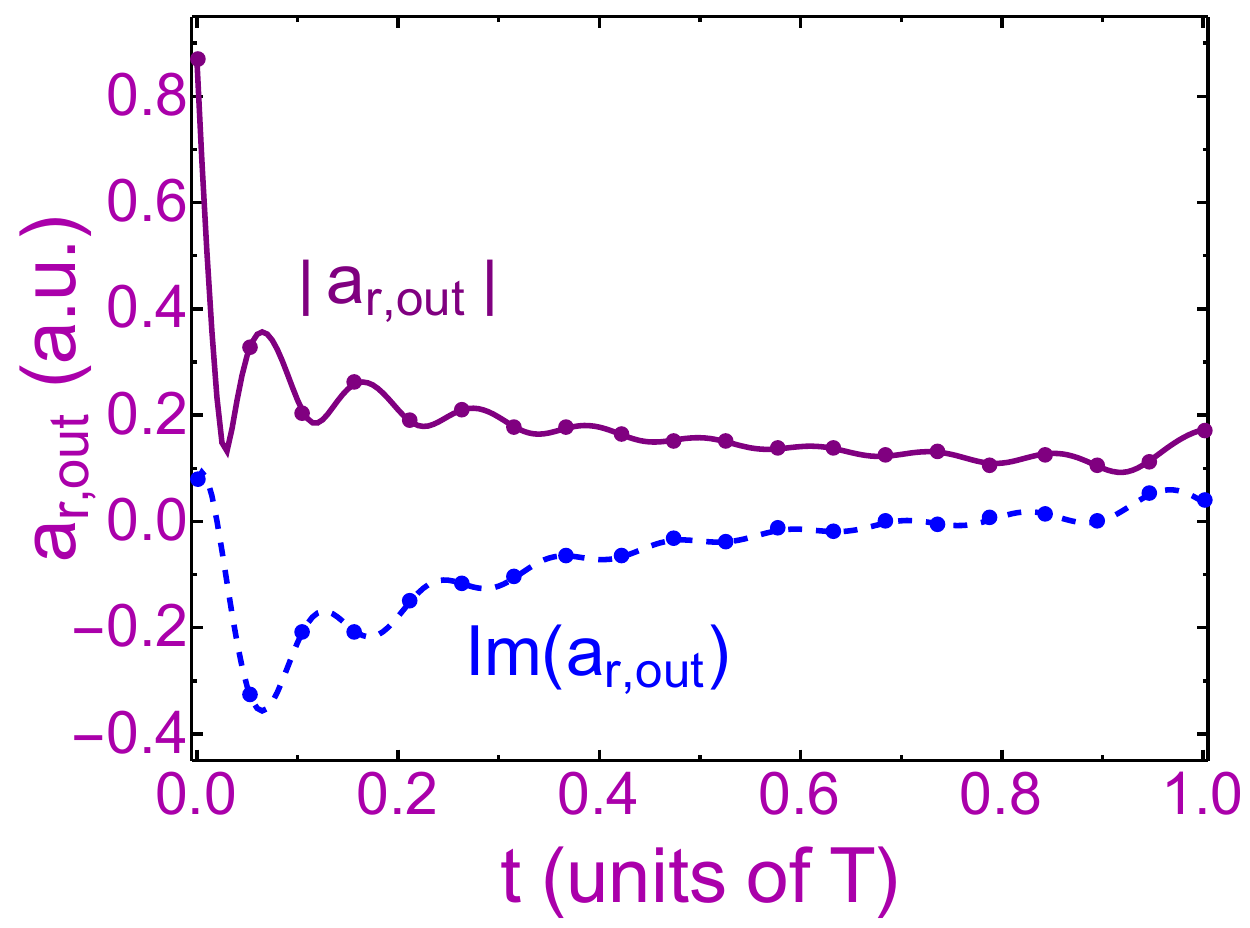, width=6.5cm}
\end{center}
\caption{(Color online) Analytical solution [dotted, based on Eq. (\ref{aRetrieval_ExpDecay})] and numerical simulation (solid and dashed lines) for the output gradient echo $a_\text{r,out}(t)$ of a quasi-monochromatic input photon wave package $a_{\text{in}}(t)=e^{-\gamma (t-t_{\text{in}})}$, $t_{\text{in}} = -T/2$. Parameters are taken to be the same as in Fig. \ref{FigStoreExpDecay}, with $\omega_m^\prime = -\omega_m = -\beta L/2$. }\label{FigRetrieveExpDecay}
\end{figure}

Next let us consider the same examples as in Sec. \ref{GEM_sub1} for retrieval process. From Eq. (\ref{ar3}), for the infinite broad-band input field described by Eq. (\ref{DeltaInput}), the gradient echo is
\begin{align}
& a_\text{r,out}(t)=-\mu \beta Le^{i\beta \frac{L}{2}\left( t+t_{\text{in}
}\right) }e^{i(\omega _{m}^{\prime }t-\omega _{m}t_{\text{in}})} e^{-\gamma (t-t_{\text{in}})} \times \notag \\
& \Phi _{2}\left( i\mu +1,-2i\mu ;2;-i\beta L\left( t+t_{\text{in}}\right) ,-i\beta Lt\right).
\end{align}
On the other hand, for qusimonochromatic narrow-band input field described by Eq. (\ref{ExpDecayInput}), in the case of nonvanishing frequency shift of GEM given by Eq. (\ref{ExpDecayInputFreqShift}), the exact solution of the output gradient echo is (see Appendix \ref{AppSec_Retrieval})
\begin{align}
a_\text{r,out}(& t)=e^{-\gamma (t-t_{\text{in}})}e^{i\left( \beta \frac{L}{%
2}+\omega _{m}^{\prime }\right) t} \times \notag \\
&  \big[ \Phi _{2}(i\mu ,-2i\mu;1;-i\beta L\left( t-T\right) ,-i\beta Lt)- \notag \\
& \qquad \qquad \qquad \Hypergeometric{1}{1}{-i\mu}{1}{-i\beta L t}\big], \label{aRetrieval_ExpDecay}
\end{align}
which is plotted in Fig. \ref{FigRetrieveExpDecay}.

\section{Gradient frequency comb}
\label{GFC}
Quite recently, an interesting quantum memory scheme---gradient frequency comb (GFC)---was proposed for the storage of a single-photon wave packet. Unlike AFC~\cite{Afzelius09}, the GFC comb teeth are distributed along the incoming photon propagation direction rather than being together uniformly everywhere in the medium. The mechanism of GFC scheme is the same as AFC, and the single-photon storage and processing techniques developed in AFC, such as on-demand retrieval~\cite{Afzelius09, Afzelius10,Lauritzen11,Akhmedzhanov13}, temporal sequencing~\cite{Saglamyurek14}, time-bin qubit measurement~\cite{Usmani10}, etc, are all applicable in GFC scheme.
Beyond this, by assigning different comb teeth to different positions, the preparation of the frequency comb is simplified, leading to various ways for the implementation of this scheme.
For example, GFC is proposed to be realized in M\"ossbauer solids using Doppler-shifted nuclear transitions for $\gamma$-ray quantum memory~\cite{Zhang15TBPgamma}, and in atomic ensembles using $\Lambda$-level structures based on off-resonant Raman configuration for optical photon storage~\cite{Zhang15TBP}.

\begin{figure}[h]
\begin{center}
\epsfig{figure=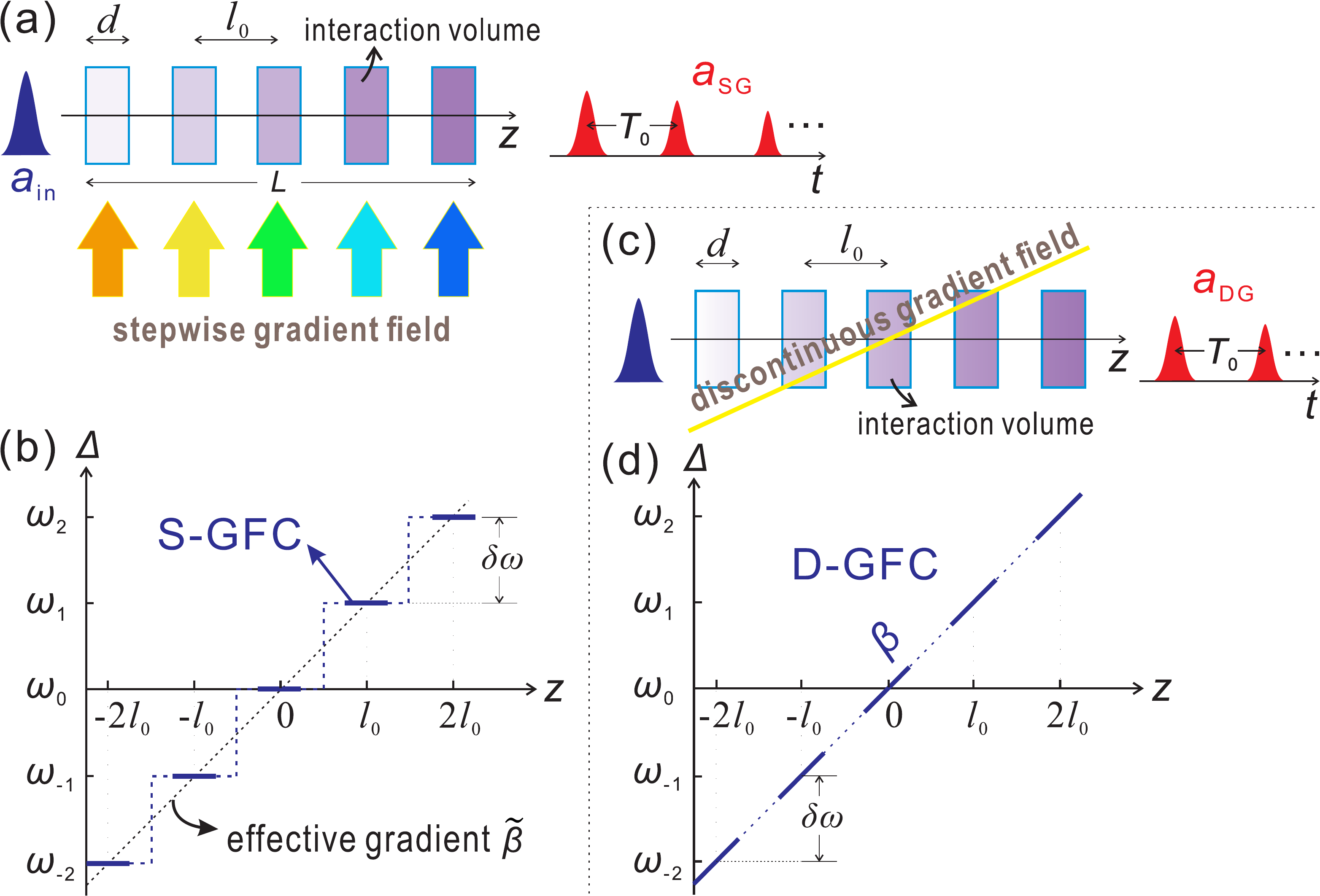, width=7.88cm}
\end{center}
\caption{(Color online) Illustration of (a) stepwise and (b) discontinuous gradient frequency comb.}\label{FigGFC}
\end{figure}

In all previous examples of GFC~\cite{Zhang15TBPgamma, Zhang15TBP}, the input photon detuning is a stepwise function of space [see Fig. \ref{FigGFC} (b)]. For this reason, we call it stepwise-gradient frequency comb (S-GFC), where the gradient is an effective gradient manifesting only when viewing its comb teeth all together. Here we suggest and analysis another alternative: discontinuous-gradient frequency comb (D-GFC), in which the photon detuning is discontinuous in space with an otherwise true gradient on it, as shown in Fig. \ref{FigGFC} (d). In other words, each comb tooth has an internal gradient built inside. We are interested in how this internal gradient modifies the echoes compared with S-GFC.

Specifically, we propose a simple version of GFC, which is realized by applying electric or magnetic field with longitudinal gradient (or effective gradient in the case of S-GFC) to a two-level medium with discrete interaction volume. Due to Stark or Zeeman effect, different sections of the interaction volume have different transition frequencies, whose adjacent spectral spacings are kept the same to form a regular comb.
Such spectral spacing between neighboring comb teeth can be adjusted by additional dc bias field [Fig. \ref{FigGFC} (a)] and/or the physical spacing between different sections [Fig. \ref{FigGFC} (c)]. So, our comb is generated out of a gradient frequency by simply breaking the continuity of GEM scheme.

As shown in Fig. \ref{FigGFC} (c), a single-photon (or weak field) wave packet of FWHM duration $\Delta t$ interacts with a medium of decoherence rate $\gamma$ subjecting to a frequency gradient $\beta$. The interaction volume of the sample is composed of $M$ discrete parts, each corresponding to one GFC comb tooth. The $m^\text{th}$ segment of the sample ($m=0, \pm 1, \pm 2, \cdots, \pm M_0$) has a central position $l_m$ and medium length $d$. In D-GFC, without loss of generality, we can assume there is no bias field and $\l_m=m l_0$, where $l_0 \geqslant d$ is the distance between the centers of two neighboring segments. The spectral separation between two adjacent comb teeth is $\delta \omega = \beta l_0 $, which gives rephasing time between two echoes $T_0=2\pi / \delta \omega$.
For the sake of simplicity, in Sec. \ref{GFC_sub2}, in the discussion of D-GFC we assume $\gamma \ll \beta d$. This means that the spectral width of each comb tooth is $\beta d$, therefore the comb finesse is defined as $\mathcal{F}^{\prime }=l_{0}/d$. We also introduce $\mu = |g|^2N/\beta$, where $g$ is the field-atom coupling constant and $N$ is the atomic density.

As shown in Fig. \ref{FigGFC} (a), the S-GFC scheme follows similarly, with comb teeth spacing $\delta \omega$ determined by the dc bias field. Since the spectral width of each S-GFC comb teeth is $2\gamma$, its finesse is defined as $\mathcal{F}=\delta \omega/(2\gamma)$. In S-GFC, for the purpose of comparison, it is useful to introduce an effective gradient $\tilde{\beta}=BW/L$ and an effective spacing $l_{0}$ such that $\delta \omega = \tilde{\beta} l_{0}$, where $BW$ is the total storage bandwidth of the comb and $L$ is the total medium length including the the free space between neighboring sections. With these two definitions we can introduce in S-GFC regime $\mu =|g|^{2}N/\tilde{\beta}$ and $\mathcal{F}^{\prime }=l_{0}/d$ just in the same way as for D-GFC.
We will use the subscript ``SG" to denote a function in the regime of ``stepwise gradient" frequency comb, and ``DG" to denote ``discontinuous gradient" frequency comb.

Similar to Eqs. (\ref{GEMa}) and (\ref{GEMS}), the evolution equation of GFC scheme can be written as:
\begin{align}
\frac{\partial }{\partial z}a(z,t)& =g^{\ast}N\sum_{m=-M_{0}}^{M_{0}}S^{(m)}(z,t)\Theta ^{(m)}(z),  \label{EQDGa1} \\
\frac{\partial }{\partial t}S^{(m)}(z,t)& =-\left[ \gamma -i\Delta ^{(m)}(z)\right] S^{(m)}(z,t)-ga(z,t),  \label{EQDGS1}
\end{align}
where
\begin{align}
\Delta ^{(m)}(z)& =(m\delta \omega^\prime +\beta z)\Theta ^{(m)}(z), \\
\Theta ^{(m)}(z)& =\Theta \left( z-l_{m}+\frac{d}{2}\right) -\Theta \left(z-l_{m}-\frac{d}{2}\right) ,
\end{align}
$z\in \lbrack -L/2,L/2]$. For D-GFC, the bias $\delta \omega ^\prime$ (if any) can be incorporated into the definition of $l_0$, so we will simply take $\delta \omega ^\prime=0$, as discussed above. For S-GFC, the gradient inside each comb teeth is $\beta=0$, so the comb teeth spacing is $\delta \omega = \delta \omega^\prime = \tilde{\beta}l_0$.

GFC is a hybrid of AFC and GEM. But unlike GEM, GFC does not require a reverse of the gradient of the external dc field, because the periodic beating among different frequency components of the polarization generates a series of echoes, which corresponds to an on-pseudodemand retrieval as in AFC. The on-demand retrieval can be achieved, on the other hand, in two ways in two-level GFC. The first way is to change the rephasing time by tuning the gradient of the external field after the complete absorption of the incoming photon. This will establish a new constructive interference condition and consequently modify the emergence time of the echo. The second way is to operate the scheme in the ``stepwise gradient echo memory" regime~\cite{Zhang15TBPgamma, Zhang15TBP} by reversing the electric or magnetic field gradient. This is a direct analogue of GEM but carried out in a discrete manner, where the first echo is generated not because of the beating, but rather due to the rephasing, of the polarization components. However, in this paper we will assume a time-independent gradient of GFC scheme, and only focus on the on-pseudodemand retrieval of GFC echo.

\subsection{GFC echoes from an optically thin medium}
\label{GFC_sub1}
The GFC scheme can be considered as a series of small, elementary GEM units aligned along $\hat{z}$, each of which has length $d$, initial condition $s(z,-\infty)=0$ and boundary condition determined by the output of the previous section. So we can apply the response function $f_s$ of GEM storage process in Eqs. (\ref{GEMstorageResp}) and (\ref{NoGRADstorageResp}) of Sec. \ref{GEM_sub1} to each section of the interaction volume to calculate GFC echoes. In D-GFC regime, each small GEM has $\beta \neq 0$, and for S-GFC $\beta = 0$. For convenience, let us denote $a(t)=a(d,t)$ and $f(t)=f_s(d,t)$. From Eqs. (\ref{as2}) and (\ref{GEMstorageResp}), we have the D-GFC field
\begin{equation}
a_{\text{DG}}^{(m)}(t)=\int_{-\infty }^{\infty }d\tau a_{\text{DG}}^{(m-1)}(t-\tau )f_{\text{DG}}^{(m)}(\tau ), \label{aDG1}
\end{equation}
where
\begin{align}
f_{\text{DG}}^{(m)}(t) = &\delta (t)- \mu \beta de^{-i\left( \beta\frac{d}{2}-\omega _{m}\right) t} \times \notag \\
& e^{-\gamma t}\,\Hypergeometric{1}{1}{i\mu +1}{2}{i\beta d t}\Theta (t). \label{fDG1}
\end{align}
The additional frequency shift introduced in Sec. \ref{GEM} now has the meaning of the central frequency of $m^\text{th}$ comb tooth: $\omega_m = m \delta \omega$, $\delta \omega > 0$.

\begin{figure}[h]
\begin{center}
\epsfig{figure=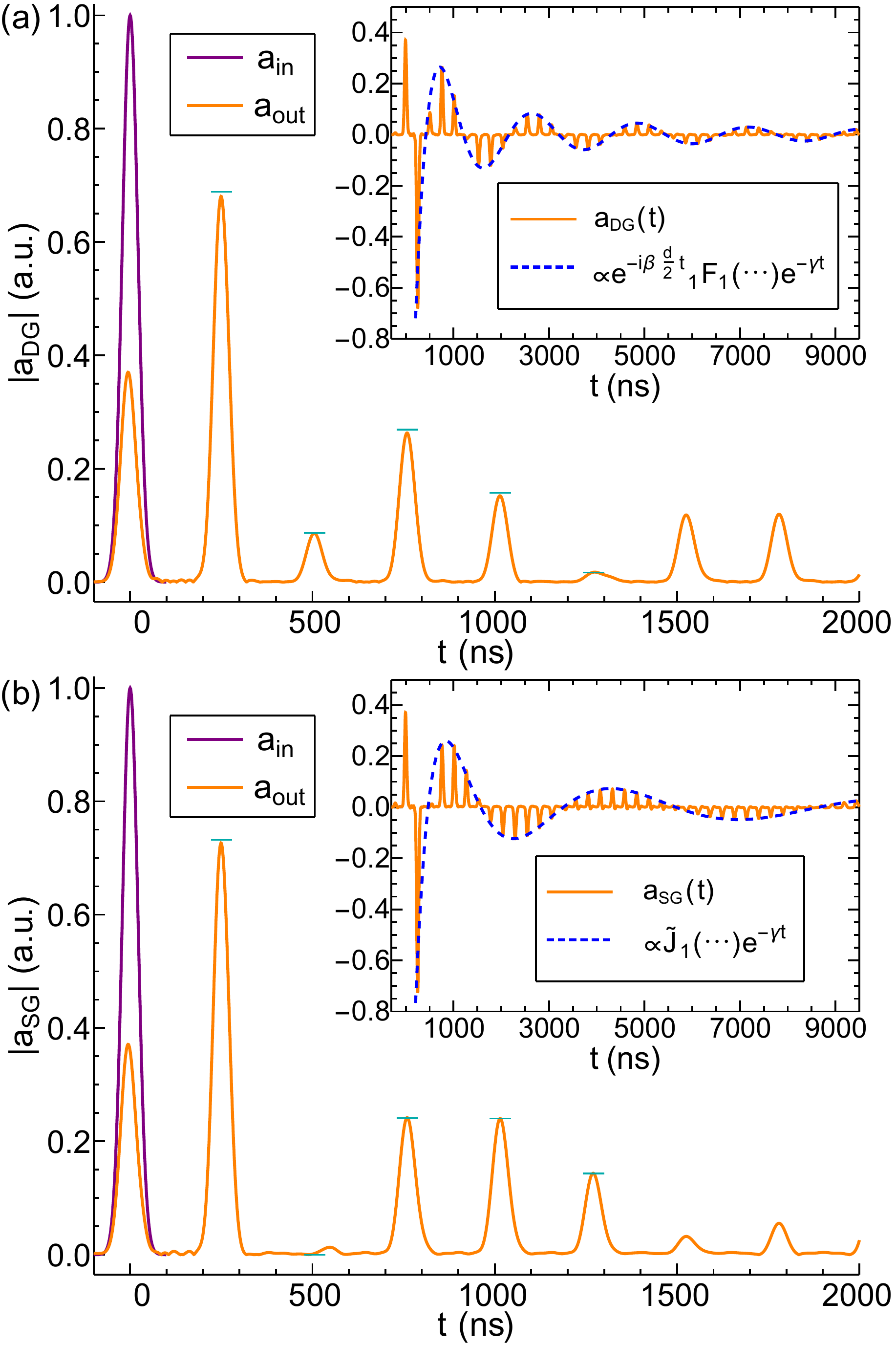, width=8cm}
\end{center}
\caption{(Color online) D-GFC (a) and S-GFC (b) echoes as functions of time for $|g|^2 NdT_0 = 2$. The solid lines are numerical simulations of Eqs. (\ref{EQDGa1}) and (\ref{EQDGS1}). The insets show that in this case the higher sequence of echoes are described by Eqs. (\ref{aDG2}) and (\ref{aSG2}) for D- and S-GFC respectively. The solid bars on the first five echoes are the predicted amplitudes of the echoes given by Eqs. (\ref{App_DG_firstFive}) and (\ref{App_SG_firstFive}). The parameters are as follows: $\Delta t=50$~ns, $M=11$, $\mathcal{F}=600$, $\mathcal{F}^\prime = 5$, $\mu = 5/\pi$, $|g|^2 N = 4 \times 10^{10}$~m$^{-1}$s$^{-1}$, $\beta L = \tilde{\beta} L = 4\pi / \Delta t$, $T_0 = 255$~ns.}\label{FigGFCopticalThin}
\end{figure}
The recursion of Eq. (\ref{aDG1}) from $m=-M_0$ to $M_0$ gives the D-GFC echoes in an optically thin medium. Similarly, the S-GFC echoes can also be calculated in the same way. The results are shown in Appendix \ref{AppSec_GFC1_sub1} and \ref{AppSec_GFC1_sub2} respectively, which read:
\begin{align}
a_{\text{DG}}^{(M_{0})}(t)\approx & a_{\text{lk}}(t)- |g|^{2}NdT_0 \sum_{n=1}^{\infty }a_{\text{in}}( t- nT_0) \times \notag \\
e^{-\gamma n T_0 } & e^{-i\beta \frac{d}{2} n T_0}\Hypergeometric{1}{1}{i\mu +1}{2}{i\beta d nT_0},  \label{aDG2}
\end{align}
\begin{align}
a_{\text{SG}}^{(M_{0})}(t)\approx & a_{\text{lk}}(t) - |g|^{2}NdT_0 \sum_{n=1}^{\infty }a_{\text{in}}( t- nT_0) \times \notag \\
& e^{-\gamma n T_0 }\tilde{J}_{1}(|g|^{2}Nd n T_0),  \label{aSG2}
\end{align}
where $a_{\text{lk}}(t)$ is the leakage field that is not absorbed by the medium
\begin{align}
a_{\text{lk}}(t) = \left( 1 - \frac{1}{2}|g|^{2}NdT_0 \right)a_{\text{in}}(t).
\end{align}
In the insets of Fig. \ref{FigGFCopticalThin} we plot analytical solutions (\ref{aDG2}) and (\ref{aSG2}), and show that they agree with the numerical simulation.

Equations (\ref{aDG2}) and (\ref{aSG2}) are valid for a narrow-band input field $2\pi / \Delta t <M\delta \omega $ interacting with an optically thin medium. Notice that in this section we specify an ``optically thin medium" by the condition $\zeta^0_\text{eff} \lesssim 2/\pi$, where $\zeta^0_\text{eff}=(2/\pi)|g|^{2}Nd T_0$ is an ``individual effective optical thickness" equal to the optical thickness of each section of the medium $\zeta^0 = 2|g|^2Nd/\gamma$ divided by the comb finesse $\mathcal{F}=\delta \omega / (2\gamma)$. The narrow-band input condition $2\pi / \Delta t <M\delta \omega $ allows us to take the approximation $M\rightarrow \infty $ when summing over the contributions from $M$ elementary GEM units. The optically thin medium condition $|g|^{2}Nd T_0 \lesssim 1$, on the other hand, allows us to truncate at the process in which each section of the medium only interacts with the original input field. This corresponds to a first order approximation, where we neglect all the re-emission and re-absorption processes.
Here the ``re-emission" and ``re-absorption" are used to refer to the processes \emph{in space} happening among discrete sections of the medium, i.e., differen elementary GEM units. As mentioned in Sec. \ref{GEM_sub1}, the second term of Eq. (\ref{fDG1}) is the response describing the part of the incoming field being absorbed.
In other words, it is the field released into space by the $m^\text{th}$ GEM unit with an opposite phase compared with the input. This part of the field will be likely absorbed and emitted again by the subsequent units. But when the effective optical thickness of each GEM unit is not high, i.e., $|g|^{2}Nd T_0 \lesssim 1$, such crosstalks between different sections of the medium is insignificant, so that once the phases of the polarizations get back together after integer multiples of $T_0$, the system resumes its evolvement following the \textit{elementary} GEM response [noticing that the argument in the response is $|g|^2Ndt$ instead of $|g|^2N(Md)t$].
That is to say, in such a case, the higher sequence of echoes subject to a phase and amplitude modulation determined by functions $e^{-i\beta \frac{d}{2} n T_0}\Hypergeometric{1}{1}{i\mu +1}{2}{i\beta d nT_0}$ and $\tilde{J}_{1}(|g|^{2}Nd n T_0)$ in D- and S-GFC respectively, following a similar manner as in the usual continuous GEM.

However, in an optically dense medium, $|g|^{2}Nd T_0 \gg 1$, the output of field from previous divisions of the medium interacts strongly with the successive divisions, so the re-emission and re-absorption processes among different GEM units dominate. In this regime, the discreteness of GEM greatly affects the field evolution such that Eqs. (\ref{aDG2}) and (\ref{aSG2}) are not valid anymore, as shown in the insets of Fig. \ref{FigGFCopticalThick}. So in order to study GFC in a more general situation, we will adopt Fourier transformation method of $t$ in Sec. \ref{GFC_sub2} to calculate the first several echoes.

\subsection{First few GFC echoes}
\label{GFC_sub2}
In practice, from the optimization point of view, we need to know the solution of a specific GFC echo valid for the full range of parameter space, rather than an expression of the whole echo series applicable only for a limited optical thickness. In order to explore the complete range of optical thickness, by the method presented in Sec. \ref{GFC_sub1} (taking the $M$-fold convolution integral), we have to evaluate all the higher order processes where the re-emitted field from previous sections takes part in the interaction with the subsequent sections, which becomes mathematically too complicated. Here instead, we use the Fourier transformation method to calculate the first few echoes, which is valid in a medium of more general optical thickness.

In Appendix \ref{AppSec_GFC2} we derive the expression for the leakage field and the first five echoes [see Eqs. (\ref{App_DG_firstFive}) and (\ref{App_SG_firstFive})] by solving Eqs. (\ref{EQDGa1}) and (\ref{EQDGS1}) using Fourier transformation on $t$. The results up to the first echo are summarized as:
\begin{align}
a_{\text{DG}}(t)& = e^{-\frac{\pi \mu }{\mathcal{F}^{\prime }}} a_\text{in}(t) - 2\mu \sin \frac{\pi }{\mathcal{F}^{\prime }}e^{-\frac{\pi \mu }{\mathcal{F}^{\prime }}}a_{\text{in}}(t-T_{0}), \label{aDG01} \\
a_{\text{SG}}(t)& =e^{-\frac{\pi \mu }{\mathcal{F}^{\prime }}} a_\text{in}(t) -2\mu \frac{\pi }{\mathcal{F}^{\prime }}e^{-\frac{\pi\mu }{\mathcal{F}^{\prime }}}e^{-\frac{\pi }{\mathcal{F}}}a_{\text{in}}(t-T_{0}), \label{aSG01}
\end{align}
where $\mu = |g|^2N/\beta$ and $\mathcal{F}^{\prime}=l_0/d$.
Notice that since we are mainly interested here in the first echo optimization, for the sake of simplicity, in D-GFC regime we assume $\gamma \rightarrow 0$ in Eq. (\ref{aDG01}), which corresponds to high S-GFC finesse $\mathcal{F}\rightarrow \infty$. In such a case, the only difference between Eqs. (\ref{aDG01}) and (\ref{aSG01}) lies in $\sin (\pi /\mathcal{F}^{\prime })$ and $\pi /\mathcal{F}^{\prime }$. When $\mathcal{F}^{\prime} \gg \pi$, Eq. (\ref{aDG01}) reduces to Eq. (\ref{aSG01}). This is easy to understand because in such a case $l_{0}\gg d$, which means each comb teeth becomes spectrally so narrow that the gradient inside the comb tooth can be neglected.
\begin{figure}[h]
\begin{center}
\epsfig{figure=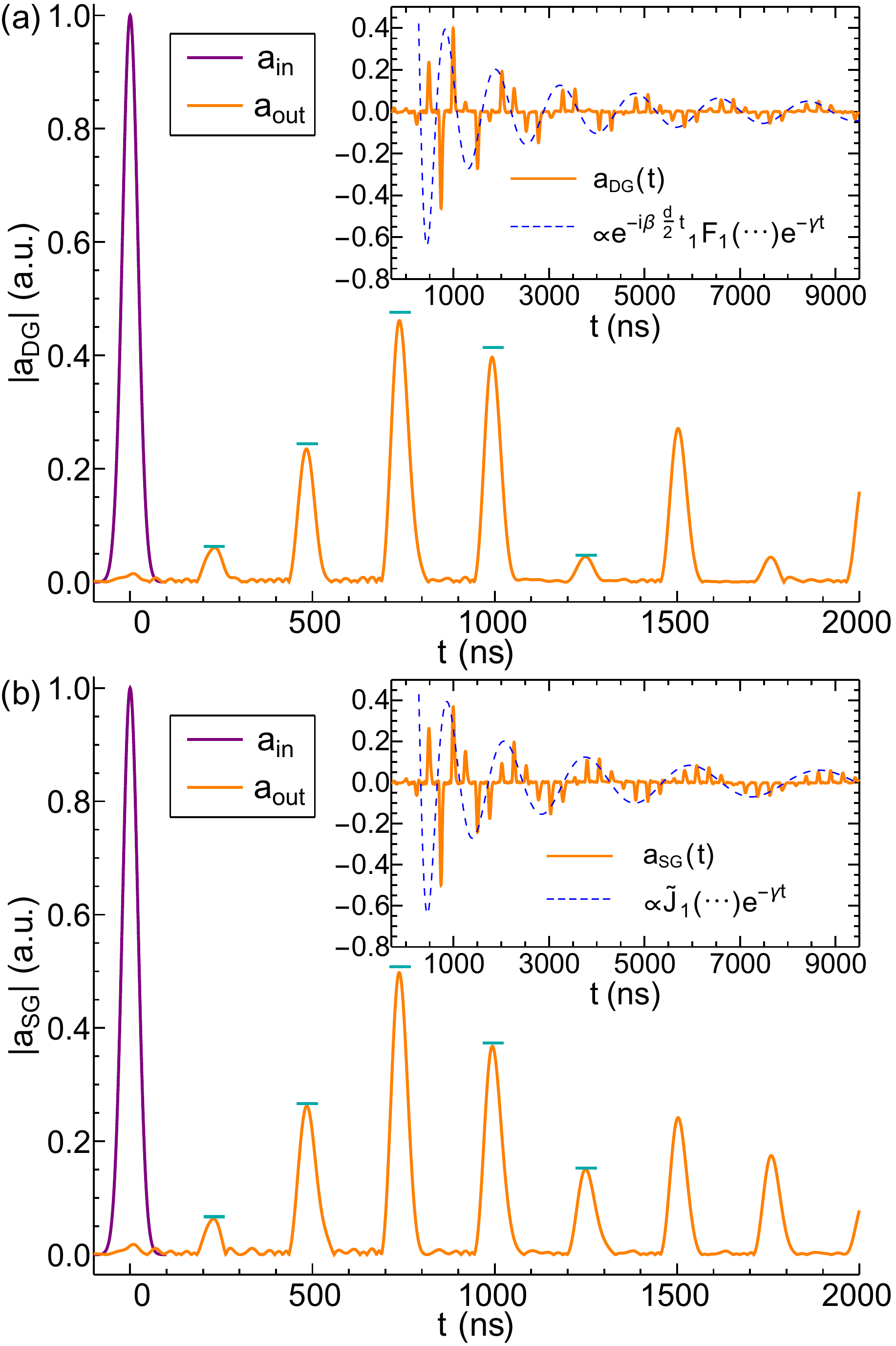, width=8cm}
\end{center}
\caption{(Color online) D-GFC (a) and S-GFC (b) echoes as functions of time for $|g|^2 NdT_0 = 10$. The solid lines are numerical simulations of Eqs. (\ref{EQDGa1}) and (\ref{EQDGS1}). The solid bars on the first five echoes are the predicted amplitudes of the echoes given by the analytical solutions Eqs. (\ref{App_DG_firstFive}) and (\ref{App_SG_firstFive}). The insets show that Eqs. (\ref{aDG2}) and (\ref{aSG2}) hardly provide useful result in this regime. However, Eqs. (\ref{App_DG_firstFive}) and (\ref{App_SG_firstFive}) [and Eqs. (\ref{aDG01}) and (\ref{aSG01})] are still valid, as shown in the main panels. The parameters are as follows: $\Delta t=50$~ns, $M=11$, $\mathcal{F}=600$, $\mathcal{F}^\prime = 5$, $\mu = 25/\pi$, $|g|^2 N = 2 \times 10^{11}$~m$^{-1}$s$^{-1}$, $\beta L = \tilde{\beta} L  = 4\pi / \Delta t$, $T_0 = 255$~ns.}\label{FigGFCopticalThick}
\end{figure}

In Figs. \ref{FigGFCopticalThin} and \ref{FigGFCopticalThick} we plot the analytical solutions of the first five D- and S-GFC echoes based on Eqs. (\ref{App_DG_firstFive}) and (\ref{App_SG_firstFive}), and show that they agree with the numerical simulation in both optically thin ($|g|^2 NdT_0 = 2$ in Fig. \ref{FigGFCopticalThin}) and thick ($|g|^2 NdT_0 = 10$ in Fig. \ref{FigGFCopticalThick}) media. On the other hand, the recursion of Eq. (\ref{aDG1}) up to the first order interaction, i.e., Eqs. (\ref{aDG2}) and (\ref{aSG2}), can not describe the GFC echoes in the later regime because the crosstalks among different elementary GEM units are too strong to be omitted, as shown in the insets of Fig. \ref{FigGFCopticalThick}.

Next let us compare the D- and S-GFC's first echo efficiencies in the regime of high finesse $\mathcal{F} \rightarrow \infty $.
From Eq. (\ref{aSG01}), the efficiency of the first S-GFC echo is
\begin{align}
\eta _{\text{SG}}^{1}& =\left(2 \mu \frac{\pi }{\mathcal{F}^{\prime }}\right)^2 \exp \left( -2\mu \frac{\pi }{\mathcal{F}^{\prime }}\right) \label{eff_SG1}
\end{align}
for $\mathcal{F} \gg 1$. Eq. (\ref{eff_SG1}) reaches maximum value $\approx 54\%$ if $\mu \pi = \mathcal{F}^{\prime } $. So the optimization condition for the first S-GFC echo is
\begin{align}
& \mathcal{F} \gg 1, \label{optimizationCond0} \\
& \Delta t < T_0 < M\Delta t, \label{optimizationCond1} \\
& \mu = \mathcal{F}^{\prime } /\pi, \label{optimizationCond2}
\end{align}
where Eq. (\ref{optimizationCond0}) reduces the role of decoherence and Eq. (\ref{optimizationCond1}) ensures the temporal resolvability of the echo ($\Delta t < T_0$) and spectral coverage of the comb ($M\delta \omega > 2\pi / \Delta t$).

\begin{figure}[h]
\begin{center}
\epsfig{figure=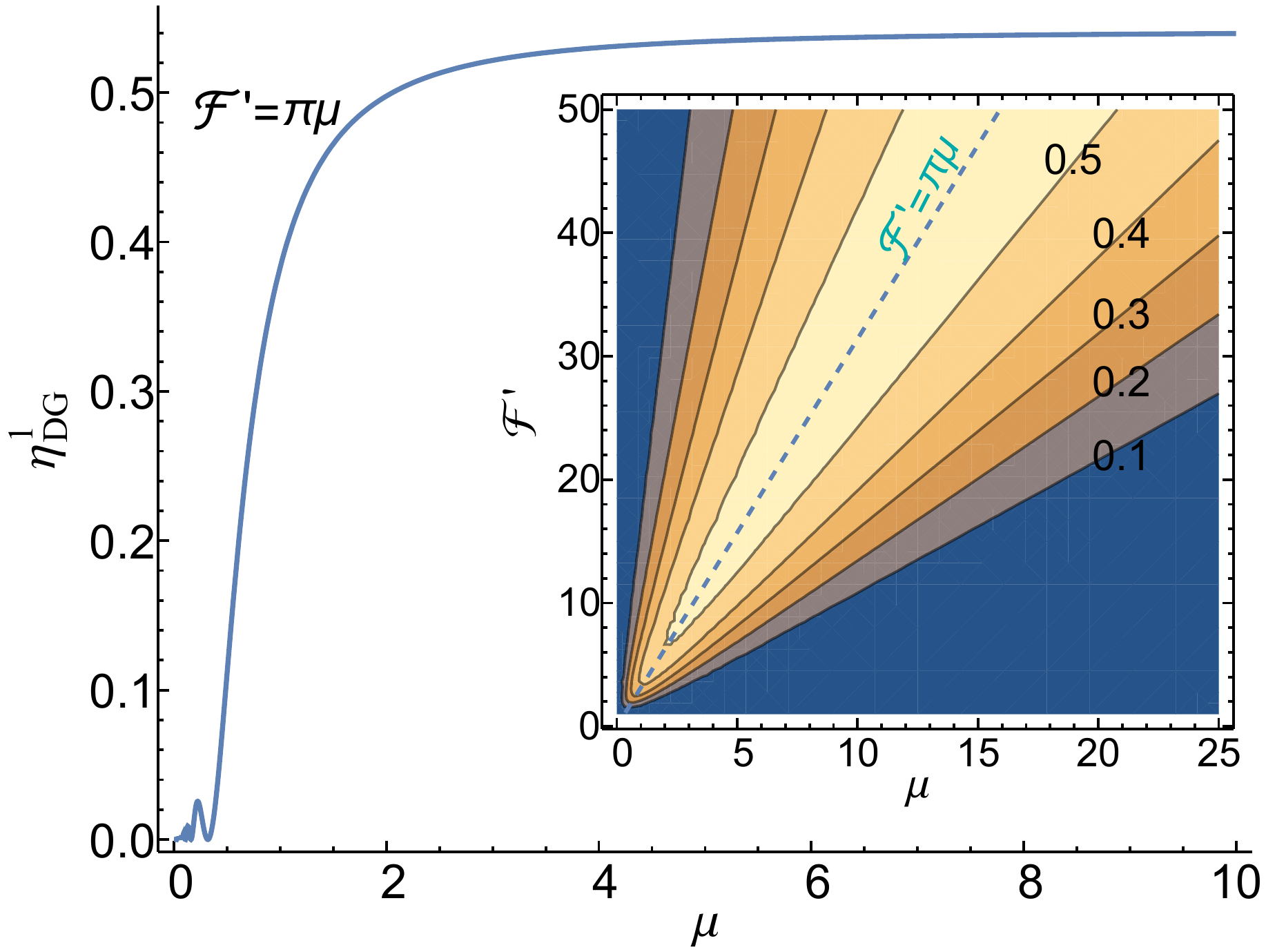, width=8cm}
\end{center}
\caption{(Color online) Analytical result of the efficiency as a function of $\mu$ for the first D-GFC echo under the optimization conditions (\ref{optimizationCond0}) - (\ref{optimizationCond2}). The inset: Contour plot of the efficiency as a function of $\mathcal{F}^{\prime }$ and $\mu$ based on Eq. (\ref{eff_DG1}). The dashed line is drawn along $\mathcal{F}^{\prime } =\pi \mu$.}\label{FigDGFCeff}
\end{figure}

From Eq. (\ref{aDG01}), the efficiency of the first D-GFC echo is
\begin{align}
\eta _{\text{DG}}^{1}& =4\mu ^{2}\sin ^{2}\frac{\pi }{\mathcal{F}^{\prime }}\exp \left( -2\mu \frac{\pi }{\mathcal{F}^{\prime }}\right). \label{eff_DG1}
\end{align}
Eq. (\ref{eff_DG1}) is maximized if Eq. (\ref{optimizationCond2}) is satisfied. The corresponding efficiency in such a case then becomes $4\mu ^{2}\sin ^{2} ( 1 /\mu ) e^{-2}$, which is still a function of $\mu$ and reaches its maximum value $\approx 54\%$ when $\mu \rightarrow \infty$ (see Fig. \ref{FigDGFCeff}). Since this expression is equal to $50\%$ when $\mu = 2.06$, the optimization conditions for the first D-GFC echo are given by Eqs. (\ref{optimizationCond0}) - (\ref{optimizationCond2}), with additional requirement
\begin{equation}
\mu \gtrsim 2. \label{optimizationCond3}
\end{equation}
Notice that in both S- and D-GFC regime, $\mu / \mathcal{F}^{\prime} = |g|^2 Nd/\delta\omega$.

So despite the differences in the two types of GFC, in the high finesse $\mathcal{F}$ regime the maximum efficiency of the first echo are both equal to $54\%$ under the optimization conditions (\ref{optimizationCond0}) - (\ref{optimizationCond2}) [also (\ref{optimizationCond3}) for D-GFC], which is similar to AFC scheme~\cite{Afzelius09, Bonarota10, Chaneliere10}. The first GFC echo is generated out of the energy that has been absorbed by the medium, and will be absorbed by the medium again when propagating through the sample. The optimization condition (\ref{optimizationCond2}) is established because of the opposite preferences on optical thickness for these two processes, i.e., to maximize the absorption of the incoming photon and minimize the reabsorption of the echo. Meanwhile, unlike GEM scheme, there is no complicated phase modulation on the GFC echoes, so the fidelity is almost $100\%$ under condition (\ref{optimizationCond1}).

\section{Discussion and conclusion}
\label{Conclusion}
Gradient echo memory (GEM) holds the promise of a quantum memory of a single-photon wave packet with high efficiency and fidelity in a forward retrieval. In this paper we derive the exact solution of a gradient echo subjecting to an arbitrary linear frequency gradient. The response function for the storage process is governed by the Kummer confluent hypergeometric function $_1F_1$, and the integral Kernel for the retrieved echo is proportional to a Humbert double hypergeometric series $\Phi_2$. This exact solution helps us to understand the physical processes of the gradient absorption of a weak field and the polarization rephasing during the formation of the echo. The rephasing process is accompanied by a phase matching procedure due to the longitudinal dependence of the transition frequency, which reduces the reabsorption of the echo and allows almost $100\%$ of the input energy to be retrieved from a medium of enough optical thickness. On the other hand, high optical thickness also smears out the rephasing and/or phase matching condition, resulting in a time-delayed and waveform-distorted echo with a strong phase modulation on it. So the optimization of GEM is reached mainly by balancing the two figures of merit, i.e., efficiency and fidelity.

By breaking the continuity of GEM we get a quantum memory scheme based on gradient frequency comb (GFC). GFC is a frequency comb with its teeth distritbuted along the photon propagation direction, therefore possessing features of both GEM and atomic frequency comb (AFC). Based on how the gradient is applied, we consider two different types of GFC: stepwise GFC and discontinuous GFC.
In the regime of high comb finesse (i.e., negligible decoherence within the rephasing time of the echo), the GFC scheme demonstrates the same efficiency and fidelity as AFC~\cite{Afzelius09, Bonarota10, Chaneliere10}.
Namely, for both of the two types of GFC, the echo fidelity can be almost $100\%$, but the efficiency of the first echo is limited to $54\%$ by the reabsorption process in the medium. Higher efficiency can be achieved by reversing the gradient of the comb, as discussed in Refs.~\cite{Zhang15TBPgamma, Zhang15TBP}.

Since the preparation of the gradient frequency comb dose not require wide inhomogeneous broadening and delicate spectral tailoring, GFC scheme can be implemented in richer varieties of materials compared with AFC. So far AFC has been realized only in rare-earth-doped crystals under cryogenic temperature, while GEM has been realized in not only rare-earth-doped materials~\cite{Hetet08EXP, Hedges10, Lauritzen10} but also warm~\cite{Hetet08EXP3level, Hosseini09, Hosseini11, Higginbottom12, Glorieux12} and cold~\cite{Sparkes13} atoms. The specific realization of GFC scheme we proposed in this paper can be carried out in these GEM systems while demonstrating the AFC performance. Meanwhile, GFC scheme is not limited to two-level system either. For example, the same D-GFC evolution equation (\ref{EQDGa1}) and (\ref{EQDGS1}) can be realized in a continuous three-level medium by discretely blocking a spatially chirped control field~\cite{Zhang14PRA} in an off-resonant Raman configuration. Considering that off-resonant Raman scheme has been implemented in cesium vapor~\cite{Reim10, Reim11, Reim12}, diamond~\cite{England13} and hydrogen~\cite{Bustard13}, a three-level GFC may extend the suitable candidate materials for AFC-like quantum memories from atomic to molecular systems. Not only that, M\"{o}ssbauer nuclear transitions have been suggested for the implementation of S-GFC scheme~\cite{Zhang15TBPgamma}, which is the only feasible quantum memory scheme in $\gamma$-ray regime so far. All these connections between GFC and other quantum memory schemes make it an important and interesting technique for the storage and retrieval of a single-photon wave packet.


\begin{acknowledgments}
I gratefully acknowledge O. Kocharovskaya, M. Scully, A. Kalachev and A. Svidzinsky for very useful discussions. The author is supported by the Herman F. Heep and Minnie Belle Heep Texas A$\&$M University Endowed Fund held/administered by the Texas A$\&$M Foundation.
\end{acknowledgments}

\appendix

\section{General solution of GEM in Laplace domain}
\label{AppSec_GeneralSln}
The equations we are solving are Eqs. (\ref{EQa2}) and (\ref{EQs2}):
\begin{align}
\frac{\partial }{\partial z}a(z,t)& =g^{\ast }Ns(z,t)e^{i\beta
tz}e^{-i\left( \beta \frac{L}{2}-\omega _{m}\right) t},  \label{App_EQa2} \\
\frac{\partial }{\partial t}s(z,t)& =-\gamma s(z,t)-ga(z,t)e^{-i\beta
tz}e^{i\left( \beta \frac{L}{2}-\omega _{m}\right) t}.  \label{App_EQs2}
\end{align}
with $t\in \lbrack -T,0]$ for storage, $t\in \lbrack 0,T]$ for retrieval, and $z\in \lbrack 0,L]$. Eqs. (\ref{App_EQa2}) and (\ref{App_EQs2}) describe the field-atom evolution of quantum memory based on GEM, PMC and control field spatial chirp with continuous frequency gradient in space. This evolution equation can be solved exactly without assumptions on the parameters, not only for storage but also for retrieval. The exact solution of Eqs. (\ref{App_EQa2}) and (\ref{App_EQs2}) for $t\in \lbrack t_{i},t_{f}]$ is derived in the following in Laplace domain of space variable $z$, with boundary condition $a(0,t)$ and initial condition $s(z,t_{i})$. Here $t_i$ ($t_f$) represents the initial (final) time with meaning depending on the context.

First let us take the spatial derivative of Eq. (\ref{App_EQs2}):
\begin{align}
\frac{\partial }{\partial z}\frac{\partial }{\partial t}s(z,t)=& -\gamma
\frac{\partial }{\partial z}s(z,t)-|g|^{2}Ns(z,t)-  \notag \\
& i\beta t\left[ \frac{\partial }{\partial t}s(z,t)+\gamma s(z,t)\right] .
\end{align}
So Eqs. (\ref{App_EQa2}) and (\ref{App_EQs2}) become
\begin{align}
& \frac{\partial }{\partial z}a(z,t)=g^{\ast }Ns(z,t)e^{i\beta
tz}e^{-i\left( \beta \frac{L}{2}-\omega _{m}\right) t},  \label{App_EQa3}
\end{align}
\begin{align}
\left( \frac{\partial }{\partial z}+i\beta t\right) \frac{\partial }{
\partial t}s(z,& t)=-\gamma \frac{\partial }{\partial z}s(z,t)-  \notag \\
& \left( |g|^{2}N+i\beta t\gamma \right) s(z,t),
\label{App_EQs3}
\end{align}
with boundary condition built in through Eq. (\ref{App_EQs2}):
\begin{equation}
\frac{\partial }{\partial t}s(0,t)=-\gamma s(0,t)-ga(0,t)e^{i\left( \beta
\frac{L}{2}-\omega _{m}\right) t}.  \label{App_BC}
\end{equation}
After taking the Laplace transformation of Eqs. (\ref{App_EQa3}) and (\ref{App_EQs3}) and substituting the boundary condition (\ref{App_BC}), we have:
\begin{align}
& pa(p,t)-a(0,t)=g^{\ast }Ns(p-i\beta t,t)e^{-i\left( \beta \frac{L}{2}
-\omega _{m}\right) t},  \label{App_EQa4p}
\end{align}
\begin{align}
& ( p+i\beta t) \frac{\partial }{\partial t}s(p,t)+\left( |g|^{2}N+ p\gamma + i\beta
t\gamma \right) s(p,t)  \notag \\
& \qquad \qquad =-ga(0,t)e^{i\left( \beta \frac{L}{2}-\omega _{m}\right) t}.
\label{App_EQs4p}
\end{align}

Solving Eq. (\ref{App_EQs4p}) with the initial condition $s(p,t_{i})$ in Laplace domain, one has
\begin{align}
s(p,t)& =s(p,t_{i})\exp \left[ -\int_{t_{i}}^{t}d\tau \left( \frac{|g|^{2}N}{
p+i\beta \tau }+\gamma \right) \right] +  \notag \\
& \int_{t_{i}}^{t}d\tau \frac{-g}{p+i\beta \tau }a(0,\tau )e^{i\left( \beta
\frac{L}{2}-\omega _{m}\right) \tau }\times  \notag \\
& \exp \left[ -\int^{t}d\tau ^{\prime \prime }\left( \frac{\left\vert
g\right\vert ^{2}N}{p+i\beta \tau ^{\prime \prime }}+\gamma \right) \right]
\times  \notag \\
& \exp \left[ \int^{\tau }d\tau ^{\prime \prime }\left( \frac{\left\vert
g\right\vert ^{2}N}{p+i\beta \tau ^{\prime \prime }}+\gamma \right) \right] .
\label{App_sp0}
\end{align}
Up to Eq. (\ref{App_sp0}), $s(p,t)$ is valid for arbitrary time dependent frequency gradient $\beta =\beta (t)$. Explicit solution is possible as
long as the inverse Laplace transformation can be calculated. In general this is difficult, but for a constant frequency gradient $\beta$ (during storage or retrieval), the exact solution can be found. In such a case, substituting the result of Eq. (\ref{App_sp0}) into Eq. (\ref{App_EQa4p}), we have the solution for
non-vanishing frequency gradient $\beta \neq 0$ given by Eqs. (\ref{App_ap}) and (\ref{App_sp}).

For a flat single-frequency absorption, the frequency gradient $\beta =0$. The corresponding evolution equation is simpler:
\begin{align}
\frac{\partial }{\partial z}a(z,t)& =g^{\ast }Ns(z,t)e^{i\omega _{m}t},
\label{App_EQab01} \\
\frac{\partial }{\partial t}s(z,t)& =-\gamma s(z,t)-ga(z,t)e^{-i\omega
_{m}t}.  \label{App_EQsb01}
\end{align}
The Laplace transformation of Eqs. (\ref{App_EQab01}) and (\ref{App_EQsb01}) are:
\begin{align}
a(p,t)& =\frac{1}{p}a(0,t)+\frac{g^{\ast }N}{p}s(p,t)e^{i\omega _{m}t},
\label{App_EQab0p1}
\end{align}
\begin{align}
\frac{\partial }{\partial t}s(p,t)& =-\gamma s(p,t)-ga(p,t)e^{-i\omega
_{m}t}.  \label{App_EQsb0p1}
\end{align}
Equations (\ref{App_EQab0p1}) and (\ref{App_EQsb0p1}) are reduced to
\begin{equation}
\frac{\partial }{\partial t}s(p,t)+\left( \gamma +\frac{|g|^{2}N}{p}\right)
s(p,t)=-\frac{1}{p}a(0,t)ge^{-i\omega _{m}t}.  \label{App_EQsb0p2}
\end{equation}
Solving for Eq. (\ref{App_EQsb0p2}) with initial condition $s(p,t_{i})$ and boundary condition $a(0,t)$, and substituting the result into Eq. (\ref
{App_EQab0p1}), we obtain the solution (\ref{App_ap_beta=0}) and (\ref{App_sp_beta=0}) of a flat single-frequency absorption.

\begin{widetext}
\begin{align}
a^{(\beta \neq 0)}(p,t)& =\frac{1}{p}a(0,t)-|g|^{2}N\int_{t_{i}}^{t}d\tau a(0,\tau
)e^{-i\left( \beta \frac{L}{2}-\omega _{m}\right) (t-\tau )}e^{-\gamma
(t-\tau )}\frac{p^{i\mu -1}}{\left[ p-i\beta \left( t-\tau \right) \right]
^{i\mu +1}}+  \notag \\
&g^{\ast }Ns(p-i\beta t,t_{i})\frac{p^{i\mu -1}}{\left[ p-i\beta \left(
t-t_{i}\right) \right] ^{i\mu }}e^{-\gamma (t-t_{i})}e^{-i\left( \beta \frac{
L}{2}-\omega _{m}\right) t},  \label{App_ap} \\
s^{(\beta \neq 0)}(p,t)& =s(p,t_{i})\left( \frac{p+i\beta t}{p+i\beta t_{i}}\right) ^{i\mu
}e^{-\gamma (t-t_{i})}-g\int_{t_{i}}^{t}d\tau a(0,\tau )e^{i\left( \beta
\frac{L}{2}-\omega _{m}\right) \tau }e^{-\gamma (t-\tau )}\frac{(p+i\beta
t)^{i\mu }}{(p+i\beta \tau )^{i\mu +1}}, \label{App_sp}
\end{align}
and for $\beta = 0$:
\begin{align}
a^{(\beta=0)}(p,t)& =\frac{1}{p}a(0,t)-|g|^{2}N\int_{t_{i}}^{t}d\tau a(0,\tau )e^{i\omega _{m}(t-\tau
)}e^{-\gamma \left( t-\tau \right) }\frac{1}{p^{2}}\exp \left[ -\frac{
|g|^{2}N}{p}\left( t-\tau \right) \right] +  \notag  \\
& g^{\ast }Ns(p,t_i)\frac{1}{p}\exp \left[ -\frac{|g|^{2}N}{p}(t-t_{i})\right] e^{-\gamma (t-t_{i})}e^{i\omega_{m}t} , \label{App_ap_beta=0} \\
s^{(\beta=0)}(p,t)& =s(p,t_i)\exp \left[ -\frac{|g|^{2}N}{p}(t-t_{i})
\right] e^{-\gamma (t-t_{i})} - g\int_{t_{i}}^{t}d\tau a(0,\tau )e^{-i\omega _{m}\tau }e^{-\gamma
\left( t-\tau \right) }\frac{1}{p}\exp \left[ -\frac{|g|^{2}N}{p}\left(
t-\tau \right) \right] , \label{App_sp_beta=0}
\end{align}
\end{widetext}
where $\mu =\left\vert g\right\vert ^{2}N /\beta $, $t\in[t_i, t_f]$.

\section{The exact analytical solution for GEM: storage}
\label{AppSec_Storage}
The time and space evolution of the field and the collective coherence during storage is given by the inverse Laplace transformation of Eqs. (\ref{App_ap}) - (\ref{App_sp_beta=0}) with $t_i=-T$ and $t_f=0$, subjecting to the initial condition $s_{s}(z,-T)=0 $ and boundary condition $a_{s}(z=0,t)=a_{\text{in}}(t)$. Here the arrival time $t_\text{in}$ of $a_{\text{in}}(t)$ is smaller than zero, and the subscript \textquotedblleft s" denotes the storage process. The spatial variable $z \in [0,L]$.

From Eqs. (\ref{App_ap}) and (\ref{App_sp}) we have the solution of a gradient absorption in Laplace domain
\begin{align}
& a_{s}^{(\beta\neq 0)}(p,t \leqslant 0)=\frac{1}{p}a_{\text{in}
}(t)-|g|^{2}N\int_{-T}^{t}d\tau a_{\text{in}}(\tau )\times  \notag \\
& \quad e^{-i\left( \beta \frac{L}{2}-\omega _{m}\right) (t-\tau
)}e^{-\gamma (t-\tau )}\frac{p^{i\mu -1}}{\left[ p-i\beta \left( t-\tau
\right) \right] ^{i\mu +1}},  \label{App_aps}
\end{align}
\begin{align}
& s_{s}^{(\beta \neq 0)}(p,t \leqslant 0)=-g\int_{-T}^{t}d\tau a_{\text{in}
}(\tau )\times  \notag \\
& \qquad \qquad e^{i\left( \beta \frac{L}{2}-\omega _{m}\right) \tau
}e^{-\gamma (t-\tau )}\frac{(p+i\beta t)^{i\mu }}{(p+i\beta \tau )^{i\mu +1}}.  \label{App_sps}
\end{align}
Similar result can be obtained for a flat single-frequency absorption out of Eqs. (\ref{App_ap_beta=0}) and (\ref{App_sp_beta=0}).

Using the inverse Laplace transformation~\cite{Prudnikov92}:
\begin{align}
& \mathfrak{L}^{-1}\left\{ (p+a)^{\lambda }(p+b)^{\nu }\right\}  \notag \\
=& \frac{e^{-az}\,\Hypergeometric{1}{1}{-\nu }{-\lambda -\nu}{(a-b)z}}{
z^{\lambda +\nu +1}\Gamma (-\lambda -\nu )} \label{App_invLaplace_1F1}
\end{align}
for $\text{Re}(\lambda +\nu )<0$, and
\begin{equation}
\mathfrak{L}^{-1}\left\{ p^{-\nu }e^{-a/p}\right\} =\left( \frac{z}{a}
\right) ^{\frac{\nu -1}{2}}J_{\nu -1}(2\sqrt{az})
\end{equation}
for $\text{Re}(\nu )>0$, where $_{1}F_{1}$ is the Kummer confluent hypergeometric function and $J_{\nu }$ is the $\nu ^{\text{th}}$ order Bessel function, we have
\begin{align}
& \mathfrak{L}^{-1}\left\{ p^{i\mu -1}\left[ p-i\beta \left( t-\tau \right)
\right] ^{-i\mu -1}\right\}  \notag \\
& \qquad \qquad \qquad =z\,\Hypergeometric{1}{1}{i\mu +1}{2}{i\beta \left( t -\tau
\right) z},  \label{App_invL1}
\end{align}
\begin{align}
& \mathfrak{L}^{-1}\left\{ \left( p+i\beta t\right) ^{i\mu }\left( p+i\beta
\tau \right) ^{-i\mu -1}\right\}  \notag \\
& \qquad \qquad =e^{-i\beta tz}\,\Hypergeometric{1}{1}{i\mu +1}{1}{i\beta
\left( t -\tau \right) z},  \label{App_invL2}
\end{align}
and
\begin{align}
& \mathfrak{L}^{-1}\left\{ \frac{1}{p^{2}}\exp \left[ -\frac{|g|^{2}N\left(
t-\tau \right) }{p}\right] \right\}  \notag \\
& \qquad \qquad \qquad =z\frac{J_{1}(2\sqrt{|g|^{2}N\left( t-\tau \right) z})
}{\sqrt{|g|^{2}N\left( t-\tau \right) z}},  \label{App_invL3}
\end{align}
\begin{align}
& \mathfrak{L}^{-1}\left\{ \frac{1}{p}\exp \left[ -\frac{|g|^{2}N\left(
t-\tau \right) }{p}\right] \right\}  \notag \\
& \qquad \qquad \qquad \,\, =J_{0}(2\sqrt{|g|^{2}N\left( t-\tau \right) z}).
\label{App_invL4}
\end{align}
Substituting Eqs. (\ref{App_invL1}), (\ref{App_invL2}) back into Eqs. (\ref{App_aps}), (\ref{App_sps}), and Eqs. (\ref{App_invL3}), (\ref{App_invL4}) back into Eqs. (\ref{App_ap_beta=0}), (\ref{App_sp_beta=0}), and making use of $\mathfrak{L}^{-1}\left\{ p^{-1}\right\} =1$, we obtain the exact solution during storage process:
\begin{widetext}
\begin{align}
a_{s}^{(\beta \neq 0)}(z,t& \leqslant 0)=a_{\text{in}}(t)-\mu \beta
z\int_{-T}^{t}d\tau a_{\text{in}}(\tau )e^{-i\left( \beta \frac{L}{2}-\omega
_{m}\right) (t-\tau )}e^{-\gamma (t-\tau )}\,\Hypergeometric{1}{1}{i\mu
+1}{2}{i\beta z (t -\tau)},  \label{App_as1} \\
s_{s}^{(\beta \neq 0)}(z,t& \leqslant 0)=-ge^{-i\beta z t}\int_{-T}^{t}d\tau
a_{\text{in}}(\tau )e^{i\left( \beta \frac{L}{2}-\omega _{m}\right) \tau
}e^{-\gamma (t-\tau )}\,\Hypergeometric{1}{1}{i\mu +1}{1}{i\beta z (t
-\tau)},  \label{App_ss1} \\
a_{s}^{(\beta =0)}(z,t& \leqslant 0)=a_\text{in}(t)-|g|^{2}Nz\int_{-T}^{t}d\tau
a_\text{in}(\tau )e^{i\omega _{m}(t-\tau )}e^{-\gamma \left( t-\tau \right) }\frac{
J_{1}(2\sqrt{|g|^{2}N z (t-\tau)})}{\sqrt{|g|^{2}N z (t-\tau)}},  \label{App_as1_beta=0} \\
s_{s}^{(\beta =0)}(z,t& \leqslant 0)=-g\int_{-T}^{t}d\tau a_\text{in}(\tau )e^{-i\omega _{m}\tau }e^{-\gamma \left( t-\tau \right) }J_{0}(2\sqrt{|g|^{2}N z (t-\tau)}). \label{App_ss1_beta=0}
\end{align}
\end{widetext}

The second term of Eq. (\ref{App_as1}) on the right hand side tells how a weak input field gets absorbed by a medium with longitudinally distributed, linear, gradient transition frequency. In Fourier domain, such absorption of the input spectrum is equal to
\begin{equation}
a_\text{in}(\omega)\left[ \left( \frac{\omega+\omega_m - \beta L/2 + i\gamma}{\omega + \omega_m + \beta (z-L/2) + i \gamma} \right) ^{i \mu} -1 \right],
\end{equation}
which can also be obtained by solving Eqs. (\ref{GEMa}) and (\ref{GEMS}) using Fourier transformation on $t$.

\begin{figure}[h]
\begin{center}
\epsfig{figure=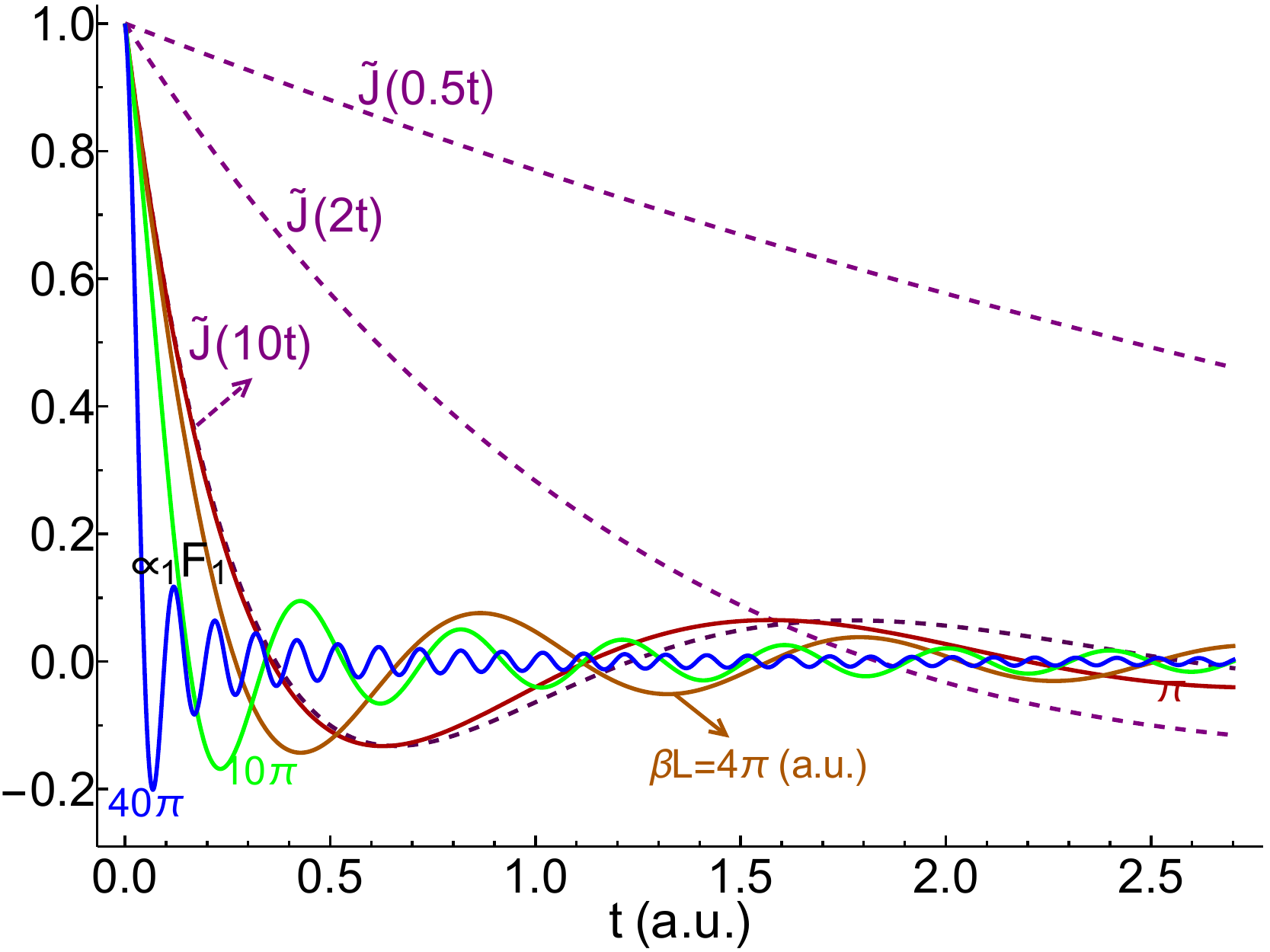, width=7.9cm}
\end{center}
\caption{(Color online) The functions of $e^{-i\beta L t/2}\,\Hypergeometric{1}{1}{i\mu+1}{2}{i\beta L t }$ (solid lines of various colors) and $\tilde{J}_{1}(|g|^{2}NLt)$ (dashed lines of purple color), where $\tilde{J}_{1}(x) = J_{1}(2\sqrt{x})/\sqrt{x}$. The solid lines are plotted under $|g|^2NL=10$~a.u., among which the red, orange, green and blue colors represent $\beta L = \pi, 4\pi, 10\pi, 40\pi $~a.u. respectively. }\label{FigResponseFunc}
\end{figure}

Comparing Eqs. (\ref{App_as1}) and (\ref{App_as1_beta=0}), it is seen that
the difference between a gradient absorption and flat single-frequency absorption lies in $e^{-i\beta tL/2}\,\Hypergeometric{1}{1}{i\mu
+1}{2}{i\beta z t }$ and $\frac{J_{1}(2\sqrt{|g|^{2}Nzt})}{\sqrt{|g|^{2}Nzt}}$. In Fig. \ref{FigResponseFunc} we plot these two functions for different parameters.

Using
\begin{align}
& \Hypergeometric{1}{1}{\alpha}{\nu}{x}=\sum_{n=0}^{\infty }\frac{\left(
\alpha \right) _{n}}{(\nu )_{n}}\frac{x^{n}}{n!}, \label{App_1F1_expension}
\end{align}
\begin{align}
& J_{1}(x)=\frac{x}{2}\sum_{n=0}^{\infty }\frac{(-1)^{n}x^{2n}}{(n+1)!n!2^{2n}},
\end{align}
where $(\alpha )_{n}=\alpha (\alpha +1)(\alpha +2)\cdots (\alpha +n-1)=\Gamma (\alpha +n)/\Gamma (\alpha )$, $(\alpha )_{0}=1$, we can write the expansions (\ref{1F1expand}) and (\ref{J1expand}), which show that the gradient absorption (\ref{App_as1}) indeed reduces to the single-frequency absorption (\ref{App_as1_beta=0}) in the limit $\beta \rightarrow 0$. This is confirmed by the following property of the Kummer confluent hypergeometric function~\cite{Magnus53}:
\begin{equation}
\lim_{\alpha \rightarrow \infty }\Hypergeometric{1}{1}{\alpha}{\nu}{-\frac{x}{\alpha}}=\Gamma (\nu )x^{\frac{1-\nu }{2}}J_{\nu -1}(2\sqrt{x}).
\end{equation}

In the small gradient regime ($\beta \ll 1$), to the first order approximation,
\begin{align}
& \left( 1-i\frac{1}{\mu }\right) \left( 1-i\frac{2}{\mu }\right) \cdots
\left( 1-i\frac{n}{\mu }\right)  \notag \\
\approx & 1-\sum_{m=1}^{n}i\frac{m}{\mu }=1-\frac{i}{2\mu }n(n+1).
\label{App_betaExpand}
\end{align}%
So from Eqs. (\ref{1F1expand}), (\ref{J1expand}) and (\ref{App_betaExpand}), we get the
first order expansion of $\Hypergeometric{1}{1}{i\mu + 1}{2}{i\beta z t }$ in terms of $\frac{J_{1}(2\sqrt{|g|^{2}Nzt})}{\sqrt{|g|^{2}Nzt}}$:
\begin{align}
& \Hypergeometric{1}{1}{i\mu + 1}{2}{i\beta z t}  \notag \\
\approx& \frac{J_{1}(2\sqrt{|g|^{2}Nzt})}{\sqrt{|g|^{2}Nzt}}+\frac{
i\left\vert g\right\vert ^{2}Nzt}{2\mu }\sum_{n=0}^{\infty }\frac{\left(
-\left\vert g\right\vert ^{2}Nzt\right) ^{n}}{(n+1)!n!}  \notag \\
=& \left( 1+i\frac{\beta zt}{2}\right) \frac{J_{1}(2\sqrt{|g|^{2}Nzt})}{
\sqrt{|g|^{2}Nzt}}. \label{App_F11Expend1}
\end{align}
From Eq. (\ref{App_F11Expend1}) we obtain Eq. (\ref{F11Expend2}) for $\beta \ll 1$.

In general, using the following integral representation of $\Hypergeometric{1}{1}{\alpha}{\nu}{x}$~\cite{Magnus53} :
\begin{align}
\Hypergeometric{1}{1}{\alpha}{\nu}{x}=& \frac{\Gamma (\nu )}{\Gamma (\nu
-\alpha )}e^{x}x^{(1-\nu )/2}\times  \notag \\
\int_{0}^{\infty }& d\xi e^{-\xi }\xi ^{(\nu -2\alpha -1)/2}J_{\nu -1}(2
\sqrt{x\xi }),
\end{align}
$\Hypergeometric{1}{1}{i\mu + 1}{2}{i\beta z t }e^{-i\beta \frac{L}{2}t}$ can be expanded as:
\begin{align}
& \Hypergeometric{1}{1}{i\mu + 1}{2}{i\beta t z}e^{-i\beta \frac{L}{2}t}
\notag \\
=& \frac{e^{i\beta \left( z-\frac{L}{2}\right) t}}{\Gamma (1-i\mu )}
\int_{0}^{\infty }d\xi e^{-\xi }\xi ^{-i\mu }\frac{J_{1}(2\sqrt{i\beta zt\xi
})}{\sqrt{i\beta zt\xi }}  \label{App_1F1integral1} \\
=& \frac{e^{i\beta \left( z-\frac{L}{2}\right) t}\mu ^{1-i\mu }}{\Gamma
(1-i\mu )}\int_{0}^{\infty }d\xi e^{-\mu (\xi +i\ln \xi )}\frac{J_{1}(2\sqrt{
i|g|^{2}Nzt\xi })}{\sqrt{i|g|^{2}Nzt\xi }}.  \label{App_1F1integral2}
\end{align}

Similarly, for the comparison between Eqs. (\ref{App_ss1}) and (\ref{App_ss1_beta=0}), $\Hypergeometric{1}{1}{i\mu +1}{1}{i\beta z t }$ reduces to $J_{0}(2\sqrt{|g|^{2}Nzt})$ when $\beta \rightarrow 0$, and
we have
\begin{align}
& \Hypergeometric{1}{1}{i\mu + 1}{1}{i\beta z t}  \notag \\
=& \frac{e^{i\beta zt}}{\Gamma (-i\mu )}\int_{0}^{\infty }d\xi \frac{1}{\xi }
e^{-\xi }\xi ^{-i\mu }J_{0}(2\sqrt{i\beta zt\xi }) \\
=& \frac{e^{i\beta zt}\mu ^{-i\mu }}{\Gamma (-i\mu )}\int_{0}^{\infty }d\xi
\frac{1}{\xi }e^{-\mu (\xi +i\ln \xi )}J_{0}(2\sqrt{i|g|^{2}Nzt\xi }).
\end{align}

We can consider two special cases for the application of Eqs. (\ref{App_as1}) and (\ref{App_ss1}). First, we assume a broad-band input
\begin{equation}
a_{\text{in}}(t)=\delta (t-t_{\text{in}}), \text{ } -T<t_{\text{in}}<0,
\end{equation}
where $t\in[-T,0]$.
Then Eqs. (\ref{App_as1}) and (\ref{App_ss1}) give
\begin{align}
a_{s}&(z, t \leqslant 0)=\delta (t-t_{\text{in}})-\mu \beta ze^{-i \left( \beta\frac{
L}{2}-\omega_m \right)(t-t_{\text{in}})}  \times  \notag \\
&e^{-\gamma (t-t_{\text{in}})} \,\Hypergeometric{1}{1}{i\mu +1}{2}{i\beta z ( t -t_\text{in} ) }\Theta (t-t_{
\text{in}}),
\end{align}
\begin{align}
s_{s}(z,t& \leqslant 0) =-ge^{-i\beta z t}e^{i \left( \beta \frac{L}{2} - \omega_m \right) t_{\text{in}
}}e^{-\gamma (t-t_{\text{in}})} \times \notag \\
& \Hypergeometric{1}{1}{i\mu +1}{1}{i\beta z
( t - t_\text{in} )}\Theta (t-t_{\text{in}}),
\end{align}
where $z \in [0,L]$.

Second, we consider a narrow-band input
\begin{equation}
a_{\text{in}}(t)=e^{-\gamma (t-t_{\text{in}})}, \text{ } -T<t_{\text{in}}<0,
\end{equation}
$t\in[-T,0]$, and frequency shift
\begin{equation}
\omega _{m}=\beta L/2.
\end{equation}
From Eqs. (\ref{App_as1}) and (\ref{App_ss1}), we obtain
\begin{align}
a_{s}(z,t \leqslant 0) & =e^{-\gamma (t-t_{\text{in}})}-e^{-\gamma (t-t_{\text{
in}})}\mu \beta z \times \notag \\
& \int_{-T}^{t}d\tau \,\Hypergeometric{1}{1}{i\mu+1}{2}{i\beta z ( t -\tau) },
\end{align}
\begin{align}
s_{s}(z,t \leqslant 0) & =-ge^{-i\beta z t}e^{-\gamma (t-t_{\text{in}
})} \times \notag \\
& \int_{-T}^{t}d\tau \,\Hypergeometric{1}{1}{i\mu +1}{1}{i\beta z ( t
-\tau) }.
\end{align}
Since
\begin{align}
\int_{-T}^{t}d\tau & \, \Hypergeometric{1}{1}{i\mu +1}{2}{i\beta z ( t -\tau ) } \notag \\
&=\frac{1}{\mu \beta z}\left[ 1-\,\Hypergeometric{1}{1}{i\mu }{1}{i\beta z (
t +T )}\right] ,
\end{align}
\begin{align}
\int_{-T}^{t}d\tau &  \,\Hypergeometric{1}{1}{i\mu +1}{1}{i\beta z ( t -\tau ) } \notag \\
&= (t+T)\,\Hypergeometric{1}{1}{i\mu +1}{2}{i\beta z ( t +T ) },
\end{align}
we have
\begin{align}
a_{s}(z,t& \leqslant 0)=e^{-\gamma (t-t_{\text{in}})}\,
\Hypergeometric{1}{1}{i\mu }{1}{i\beta z ( t +T )},
\end{align}
\begin{align}
s_{s}(z,t \leqslant 0)&=-ge^{-i\beta z t }e^{-\gamma (t-t_{\text{in}})}(t+T) \times \notag \\
&\,\Hypergeometric{1}{1}{i\mu +1}{2}{i\beta z ( t +T ) } ,
\end{align}
where $z \in [0,L]$.

\section{The exact analytical solution for GEM: retrieval}
\label{AppSec_Retrieval}
The exact solution of gradient echo is calculated based on Eqs. (\ref{App_ap}) and (\ref{App_sp}) in retrieval time window $t\in [ t_{i},t_{f}]=[0,T]$ with parameters $\beta ^{\prime }$, $g^{\prime }$, $N^{\prime }$, $\mu ^{\prime }$, and $\omega _{m}^{\prime }$. The boundary condition is
\begin{equation}
a_{r}(z=0,t)=0,  \label{App_BC_retrieval}
\end{equation}
where the subscript \textquotedblleft r" denotes retrieval. In three-level medium quantum memory methods, such as schemes based on PMC and/or control field spatial chirp, the initial coherence of retrieval process can be written as $s_{r}(z,0)=s_{s}(z,0)e^{i \kappa ^{\prime }z}$, where $\kappa ^{\prime }z$ is an additional position-dependent phase $-\phi_0(z)$ in Eq. (\ref{phi1}) exerted by external method, e.g., via a phase modulation of the read-out control field. This phase allows the echo to be shifted~\cite{Zhang14PRA} and sequenced~\cite{Zhang15TBP} in time without being
compressed or stretched. Here we are primarily interested in the usual GEM method, where $\kappa ^{\prime }=0$. In such a situation, from Eq. (\ref{App_sps}), we have
\begin{align}
& s_{r}(p,0)=s_{s}(p,0)  \notag \\
=& -g\int_{-T}^{0}d\tau a_{\text{in}}(\tau )e^{i\left( \beta \frac{L}{2}
-\omega _{m}\right) \tau }e^{\gamma \tau }\frac{p^{i\mu }}{(p+i\beta \tau
)^{i\mu +1}}.  \label{App_IC_retrieval}
\end{align}
Substituting the boundary and initial conditions (\ref{App_BC_retrieval}) and (\ref{App_IC_retrieval}) into Eqs. (\ref{App_ap}) and (\ref{App_sp}), we obtain the GEM echo and retrieval collective coherence in the Laplace domain:
\begin{align}
a& _{r}(p,t\geqslant 0)=-gg^{\prime \ast }N^{\prime }\int_{-T}^{0}d\tau a_{
\text{in}}(\tau )e^{-i\left( \beta ^{\prime }t-\beta \tau \right) \frac{L}{2}
}\times  \notag \\
& e^{i(\omega _{m}^{\prime }t-\omega _{m}\tau )}e^{-\gamma (t-\tau )}\frac{
p^{i\mu ^{\prime }-1}\left( p-i\beta ^{\prime }t\right) ^{i(\mu -\mu
^{\prime })}}{\left[ p-i\left( \beta ^{\prime }t-\beta \tau \right) \right]
^{i\mu +1}},  \label{App_arp}
\end{align}
\begin{align}
s_{r}(p,t\geqslant 0)=& -g\int_{-T}^{0}d\tau a_{\text{in}}(\tau )e^{i\left(
\beta \frac{L}{2}-\omega _{m}\right) \tau }\times  \notag \\
& e^{-\gamma (t-\tau )}\frac{p^{i(\mu -\mu ^{\prime })}\left( p+i\beta
^{\prime }t\right) ^{i\mu ^{\prime }}}{(p+i\beta \tau )^{i\mu +1}}.
\label{App_srp}
\end{align}
Using the inverse Laplace transformation~\cite{Prudnikov92}:
\begin{align}
& \mathfrak{L}^{-1}\left\{ p^{\lambda }(p-a)^{\sigma }(p-b)^{\nu }\right\}
\notag \\
=& \frac{\Phi _{2}\left( -\sigma ,-\nu ;-\lambda -\sigma -\nu ;az,bz\right)
}{z^{\lambda +\sigma +\nu +1}\Gamma (-\lambda -\sigma -\nu )},
\end{align}
where $\text{Re}(\lambda +\sigma +\nu) < 0$ and $\Phi _{2}$ is the Humbert double hypergeometric series, we have
\begin{align}
& \mathfrak{L}^{-1}\left\{ \frac{p^{i\mu ^{\prime }-1}\left( p-i\beta
^{\prime }t\right) ^{i(\mu -\mu ^{\prime })}}{\left[ p-i\left( \beta
^{\prime }t-\beta \tau \right) \right] ^{i\mu +1}}\right\}  \notag \\
=& z\Phi _{2}\left( i\mu +1,i\mu ^{\prime }-i\mu ;2;i\left( \beta ^{\prime
}t-\beta \tau \right) z,i\beta ^{\prime }tz\right),  \label{App_invL5}
\end{align}
and
\begin{align}
& \mathfrak{L}^{-1}\left\{ \frac{p^{i(\mu -\mu ^{\prime })}\left( p+i\beta
^{\prime }t\right) ^{i\mu ^{\prime }}}{(p+i\beta \tau )^{i\mu +1}}\right\}
\notag \\
=& \Phi _{2}\left( i\mu +1,-i\mu ^{\prime };1;-i\beta \tau z,-i\beta
^{\prime }tz\right).  \label{App_invL6}
\end{align}

Substituting Eqs. (\ref{App_invL5}) and (\ref{App_invL6}) back to the inverse Laplace transformation of Eqs. (\ref{App_arp}) and (\ref{App_srp}), we obtain the exact analytical expression of the field and atomic collective coherence during retrieval:
\begin{widetext}
\begin{align}
& a_{r}(z,t\geqslant 0) =-gg^{\prime \ast }N^{\prime }z\int_{-T}^{0}d\tau a_{\text{in}}(\tau )e^{-i\left( \beta ^{\prime }t-\beta \tau \right) \frac{L}{2}} e^{i(\omega _{m}^{\prime }t-\omega _{m}\tau )}e^{-\gamma (t-\tau )} \times \notag \\
& \qquad \qquad \qquad \qquad \qquad \qquad \qquad \qquad \qquad \qquad \quad \quad \Phi_{2}\left( i\mu +1,i\mu ^{\prime }-i\mu ;2;i\left( \beta ^{\prime }t-\beta\tau \right) z,i\beta ^{\prime }z t\right) ,  \label{App_ar1} \\
& s_{r}(z,t\geqslant 0) =-g\int_{-T}^{0}d\tau a_{\text{in}}(\tau )e^{i\left(\beta \frac{L}{2}-\omega _{m}\right) \tau }e^{-\gamma (t-\tau )}  \Phi
_{2}\left( i\mu +1,-i\mu ^{\prime };1;-i\beta z \tau,-i\beta ^{\prime} z t\right). \label{App_sr1} \\
& \text{If $g^{\prime }=g$, $N^{\prime }=N$, $\beta^{\prime }=-\beta $, and $\mu ^{\prime }=-\mu $:} \notag \\
& a_{r}(z,t\geqslant 0) =-\mu \beta z\int_{-T}^{0}d\tau a_{\text{in}}(\tau)e^{i\beta \frac{L}{2}\left( t+\tau \right) }e^{i(\omega _{m}^{\prime
}t-\omega _{m}\tau )}e^{-\gamma (t-\tau )}\Phi _{2}\left( i\mu +1,-2i\mu;2;-i\beta z\left( t+\tau \right) ,-i\beta z t\right) ,  \label{App_ar2} \\
& s_{r}(z,t\geqslant 0) =-g\int_{-T}^{0}d\tau a_{\text{in}}(\tau )e^{i\left(\beta \frac{L}{2}-\omega _{m}\right) \tau }e^{-\gamma (t-\tau )}\Phi
_{2}\left( i\mu +1,i\mu ;1;-i\beta z \tau,i\beta z t\right) ,  \label{App_sr2}
\end{align}
\end{widetext}
where $z\in [ 0,L]$, $t \in [ 0,T]$.

If $g^\prime = g$, $N^\prime = N$, $\beta ^{\prime }=\beta $, $\mu ^{\prime }=\mu $, and $\omega_m^\prime = \omega_m$, we should be able to recover the storage solution. Indeed, according to Eq. (\ref{App_Phi2_specialCase1}) in Appendix \ref{AppSec_Retrieval_sub1}, $\Phi _{2} ( i\mu +1,-i\mu ;1;-i\beta \tau z,-i\beta z t ) =e^{-i\beta z t}\Hypergeometric{1}{1}{i\mu +1}{1}{i\beta z (t-\tau )}$. Therefore in this case Eq. (\ref{App_sr1}) reduces to Eq. (\ref{App_ss1}).

If $g^{\prime }=g$, $N^{\prime }=N$, $\beta^{\prime }=-\beta $, and $\mu ^{\prime }=-\mu $, from Eqs. (\ref{App_ar1}) and (\ref{App_sr1}) we obtain the exact GEM solution Eqs. (\ref{App_ar2}) and (\ref{App_sr2}). From this solution, the GEM output echo (i.e., at $z=L$) can be expanded into the following form by using Eq. (\ref{App_Phi2_4}):
\begin{align}
a& _\text{r,out} (t\geqslant 0) = -\mu \beta L \sum_{n=0}^{\infty} \frac{(-2i\mu)_n}{(n+1)!}\frac{(-i\beta Lt)^n}{n!} \times \notag \\
&\int_{-T}^{0}d\tau a_{\text{in}}(\tau)e^{i\beta \frac{L}{2}(t+\tau)} e^{i(\omega _{m}^{\prime}t-\omega _{m}\tau )} e^{-\gamma (t-\tau )} \times \notag \\
& \quad \,\Hypergeometric{2}{1}{-n,i\mu +1}{2i\mu+1-n}{1+\frac{\tau}{t}} .  \label{App_ar3}
\end{align}

Now let us look at the two special examples considered before. From Eq. (\ref{App_ar2}), for a broad-band input
\begin{equation}
a_{\text{in}}=\delta (t-t_{\text{in}}),\text{ }-T<t_{\text{in}}<0,
\end{equation}
the field during retrieval is:
\begin{align}
& a_{r}(z,t\geqslant 0)
=-\mu \beta ze^{i\beta \frac{L}{2}\left( t+t_{\text{in}}\right)
}e^{i(\omega _{m}^{\prime }t-\omega _{m}t_{\text{in}})} \times \notag \\
& e^{-\gamma (t-t_{
\text{in}})}\Phi _{2}\left( i\mu +1,-2i\mu ;2;-i\beta z\left( t+t_{\text{in}
}\right) ,-i\beta zt\right) .
\end{align}
In the opposite limit, for a narrow-band input
\begin{equation}
a_{\text{in}}(t)=e^{-\gamma (t-t_{\text{in}})},\text{ }-T<t_{\text{in}}<0,
\end{equation}
and frequency shift on top of the gradient
\begin{equation}
\omega _{m}=\beta L / 2,
\end{equation}
the field during retrieval is
\begin{align}
a&_{r}(z,t\geqslant 0)
=-\mu \beta ze^{-\gamma (t-t_{\text{in}})}e^{i\left( \beta \frac{L}{2}
+\omega _{m}^{\prime }\right) t} \times \notag \\
&\int_{-T}^{0}d\tau \Phi _{2}\left( i\mu
+1,-2i\mu ;2;-i\beta z\left( t+\tau \right) ,-i\beta zt\right) .
\end{align}
Using the integral formula (\ref{App_Phi2_integral}) and relation (\ref{App_Phi2_specialCase2}) in Appendix \ref{AppSec_Retrieval_sub1},
we obtain
\begin{align}
a_{r}(z,& t \geqslant 0)
=e^{-\gamma (t-t_{\text{in}})}e^{i\left( \beta \frac{L}{2}+\omega
_{m}^{\prime }\right) t} \times \notag \\
& \big[ \Phi _{2}(i\mu ,-2i\mu ;1;-i\beta
z\left( t-T\right) ,-i\beta zt)- \notag \\
& \qquad \qquad \qquad \Hypergeometric{1}{1}{-i\mu}{1}{-i\beta z t}
\big]  . \label{App_aRetrieval_ExpDecay}
\end{align}

\section{Some properties for the Humbert double hypergeometric series $\Phi _{2}$}
\label{AppSec_Retrieval_sub1}
It is seen from Eq. (\ref{App_ar1}) that the GEM echo is mainly determined
by Humbert double hypergeometric function $\Phi _{2}$. Some of the remarks of this relatively complicated special function can be found in recent studies~\cite{Choi11, Manako11,
Rathie13, Choi15}. The function $\Phi _{2}(\alpha ,\alpha^{\prime };\nu;x,y)$ satisfies the following partial differential equation:
\begin{align}
&x \frac{\partial^2 \Phi_2}{\partial x^2} + y \frac{\partial^2 \Phi_2}{\partial x \partial y} + (\nu-x)\frac{\partial \Phi_2}{\partial x} - \alpha \Phi_2 = 0,  \\
&y \frac{\partial^2 \Phi_2}{\partial y^2} + x \frac{\partial^2 \Phi_2}{\partial x \partial y} + (\nu-y)\frac{\partial \Phi_2}{\partial y} - \alpha^\prime \Phi_2 = 0.
\end{align}
It has an integral representation%
\begin{align}
\Phi _{2}(\alpha ,\alpha ^{\prime };\nu &;x,y)=\frac{\Gamma (\nu )}{\Gamma
(\alpha )\Gamma (\alpha ^{\prime })\Gamma (\nu -\alpha -\alpha ^{\prime })}
\times  \notag \\
\int_{0}^{1}\int_{0}^{1} & d\xi d\eta e^{x\xi +y(1-\xi )\eta }\xi ^{\alpha
-1}\eta ^{\alpha ^{\prime }-1} \times  \notag \\
& (1-\xi )^{\nu -\alpha -1}(1-\eta )^{\nu-\alpha -\alpha ^{\prime }-1}.
\end{align}

Two useful relations in special cases between $\Phi_2$ and Kummer confluent hypergeometric function $_1 F_1$ are:
\begin{equation}
\Phi _{2}(\alpha ,\nu -\alpha ;\nu ;x,y)=e^{y}\,\Hypergeometric{1}{1}{\alpha}{
\nu}{x-y}, \label{App_Phi2_specialCase1}
\end{equation}
\begin{equation}
\Phi _{2}(\alpha , \alpha ^\prime ;\nu ;x,x)= \Hypergeometric{1}{1}{\alpha+\alpha^\prime}{
\nu}{x}. \label{App_Phi2_specialCase2}
\end{equation}

The function $\Phi _{2}(\alpha ,\alpha^{\prime };\nu;x,y)$ can also be defined through a double series as:
\begin{equation}
\Phi _{2}(\alpha ,\alpha ^{\prime };\nu ;x,y)=\sum_{m,n=0}^{\infty }\frac{
(\alpha )_{m}(\alpha ^{\prime })_{n}}{(\nu )_{m+n}m!n!}x^{m}y^{n},
\label{App_Phi2_1}
\end{equation}
where $(\alpha )_{m}=\alpha (\alpha +1)\cdots (\alpha +m-1)=\Gamma (\alpha +m)/\Gamma (\alpha )$, $(\alpha )_{0}=1$. Another expansion in terms
of Gauss hypergeometric function $_{2}F_{1}$ is~\cite{Rathie13}
\begin{align}
& \Phi _{2}(\alpha ,\alpha ^{\prime };\nu ;x,y)  \notag \\
=& \sum_{m=0}^{\infty }\frac{ (\alpha )_{m}}{(\nu )_{m}}
\Hypergeometric{2}{1}{-m,\alpha^\prime}{1-\alpha-m}{\frac{y}{x}}\frac{x^{m}}{
m!}.  \label{App_Phi2_2}
\end{align}
So in the case of usual gradient echo in Eq. (\ref{App_ar2}),
\begin{widetext}
\begin{align}
&\Phi _{2} ( i\mu +1,-2i\mu ;2;-i\beta z(t +\tau),-i\beta
z t ) \notag \\
 =&\frac{1}{\Gamma (i\mu +1)\Gamma (-2i\mu )}\sum_{m,n=0}^{
\infty }\frac{\Gamma (i\mu +1+m)\Gamma (-2i\mu +n)}{(m+n+1)!m!n!}(-i\beta z t )^{n} [-i\beta
z( t +\tau )]^{m}  \label{App_Phi2_3} \\
=& \frac{1}{\Gamma (-2i\mu)}\sum_{m=0}^{\infty }\frac{\Gamma (-2i\mu + m)}{(m+1)!}\frac{(-i\beta zt)^{m}}{m!}\,\Hypergeometric{2}{1}{-m,i\mu +1}{2i\mu+1-m}{1+\frac{\tau}{t}}.  \label{App_Phi2_4}
\end{align}
\end{widetext}
Eq. (\ref{App_Phi2_4}) provides a series to Eq. (\ref{App_ar2}) that illustrates how the gradient echo is constructed.

One should be careful when evaluating $\Phi _{2}$ using expansions (\ref{App_Phi2_3}) and/or (\ref{App_Phi2_4}), since for larger argument $\beta zT$ they become numerically unstable. Also it is useful to check Eqs. (\ref{App_Phi2_3}) and/or (\ref{App_Phi2_4}) with its special values of $\Phi _{2}$ by, for example, the following cases:
\[ \Phi _{2} = \left \{
\begin{array}{ll}
_{1}F_{1} ( -2i\mu ;2;-i\beta z t ) , & \text{ if } \tau =-t,  \\
_{1}F_{1} ( i\mu +1;2;-i\beta z\tau  ), & \text{ if } t = 0, \\
_{1}F_{1} ( -i\mu +1;2;-i\beta zt ), &\text{ if } \tau = 0,
\end{array}
\right.
\]
which can be obtained from Eqs. (\ref{App_arp}) and (\ref{App_invLaplace_1F1}), and/or Eq. (\ref{App_Phi2_specialCase2}).
Sometimes $\Phi_{2} $ can be numerically evaluated more easily with inverse Laplace transformation algorithm~\cite{Stehfest70,Valko04} directly from Eq. (\ref{App_arp}).

Next we derive an integral formula of Humbert double hypergeometric function
$\Phi _{2}$:
\begin{align}
&\int dx\Phi _{2}(\alpha ,\alpha ^{\prime };\nu
;x,y)=\sum_{m,n=0}^{\infty }\frac{(\alpha )_{m}(\alpha ^{\prime })_{n}}{(\nu
)_{m+n}m!n!}\frac{x^{m+1}}{m+1}y^{n}  \notag \\
& =\sum_{n=0}^{\infty }\sum_{m=1}^{\infty }\frac{\Gamma (\alpha +m-1)\Gamma
(\nu )(\alpha ^{\prime })_{n}}{\Gamma (\alpha )\Gamma (\nu +m+n-1)m!n!}
x^{m}y^{n} \notag \\
& =\frac{\Gamma (\alpha -1)\Gamma (\nu )}{\Gamma (\alpha )\Gamma (\nu -1)}
\sum_{n=0}^{\infty }\sum_{m=1}^{\infty }\frac{(\alpha -1)_{m}(\alpha
^{\prime })_{n}}{(\nu -1)_{m+n}m!n!}x^{m}y^{n} \notag \\
& =\frac{\nu -1}{\alpha -1} [ \Phi _{2}(\alpha -1,\alpha ^{\prime };\nu
-1;x,y)- \notag \\
& \qquad \qquad \qquad \qquad \qquad \Hypergeometric{1}{1}{\alpha^\prime}{\nu-1}{y} ] , \label{App_Phi2_integral}
\end{align}
where $\nu,\alpha \neq 1$, and we have used Eqs. (\ref{App_Phi2_1}) and (\ref{App_1F1_expension}). This result is used in deriving Eq. (\ref{aRetrieval_ExpDecay}) or (\ref{App_aRetrieval_ExpDecay}).

\section{Analytical analysis of GFC in optically thin
medium}
\label{AppSec_GFC1}
Let us calculate the GFC echo in an optically thin medium using the result for GEM storage process. In this section,  the ``optically thin medium" is specified by the condition $|g|^2NdT_0 \lesssim 1$. We consider two versions of GFC: discontinuous GFC (D-GFC) and stepwise GFC (S-GFC). In either one, the interaction volume of the medium is composed of $M$ segments, each having length $d$ and frequency offset $\omega_{m}$, where $m=0,\pm 1,\pm 2,\cdots \pm M_{0}$. Every segment can be treated as an independent\ small GEM system with $z\in [ 0,d]$, and the medium sees the output field from the previous section as its input:
\begin{align}
& a_{\text{in}}^{(m)}(0,t)=a^{(m-1)}(d,t),\text{ if }m\neq -M_{0}, \\
& a_{\text{in}}^{(-M_{0})}(0,t)=a_{\text{in}}(t).
\end{align}
The goal is to calculate the final output $a^{(M_{0})}(d,t)$ subjecting to the initial condition of zero collective coherence.

\subsection{Discontinuous gradient frequency comb}
\label{AppSec_GFC1_sub1}
From Eq. (\ref{App_as1}), we have
\begin{equation}
a_{\text{DG}}^{(m)}(t)=\int_{-\infty }^{\infty }d\tau a_{\text{DG}
}^{(m-1)}(t-\tau )f_{\text{DG}}^{(m)}(\tau ),  \label{App_aDG1}
\end{equation}
where the subscript ``DG" denotes ``discontinuous gradient", and
\begin{align}
f_{\text{DG}}^{(m)}(t) =&\delta (t)-\mu \beta de^{-i\left( \beta \frac{d}{2}
-\omega _{m}\right) t}\times \notag \\
& e^{-\gamma t}\,\Hypergeometric{1}{1}{i\mu +1}{2}{i\beta d t}\Theta (t).
\end{align}
From Eq. (\ref{App_aDG1}), we have
\begin{align}
a_{\text{DG}}^{(-M_{0})}(t)& =\int_{-\infty }^{\infty }d\tau _{0}a_{\text{in}
}(t-\tau _{0})f_{\text{DG}}^{(-M_{0})}(\tau _{0}),
\end{align}
\begin{align}
a_{\text{DG}}^{(-M_{0}+1)}(t)& =\int_{-\infty }^{\infty }d\tau _{1}a_{\text{
DG}}^{(-M_{0})}(t-\tau _{1})f_{\text{DG}}^{(-M_{0}+1)}(\tau _{1}),  \\
\vdots &   \notag
\end{align}
So the recursion of $a_{\text{DG}}^{(m)}(t)$ gives
\begin{align}
& a_{\text{DG}}^{(M_{0})}(t)  \notag \\
=& \int_{-\infty }^{\infty }d\tau _{0}\cdots d\tau _{2M_{0}}a_{\text{in}
}(t-\tau _{0}-\cdots -\tau _{2M_{0}})\times  \notag \\
& \qquad \qquad \qquad f_{\text{DG}}^{(-M_{0})}(\tau _{0})\cdots f_{\text{DG}
}^{(M_{0})}(\tau _{2M_{0}})  \notag \\
=& \int_{-\infty }^{\infty }d\tau _{0}\cdots d\tau _{2M_{0}}a_{\text{in}
}(t-\tau _{0}-\cdots -\tau _{2M_{0}})\times  \notag \\
& \,\delta (\tau _{0})\cdots \delta (\tau _{2M_{0}})-\mu \beta
d\int_{-\infty }^{\infty }d\tau a_{\text{in}}(t-\tau )e^{-\gamma \tau }\times
\notag \\
& \Hypergeometric{1}{1}{i\mu +1}{2}{i\beta d \tau}\Theta (\tau
)\sum_{m=-M_{0}}^{M_{0}}e^{-i\left( \beta \frac{d}{2}-\omega _{m}\right)
\tau }+  \notag \\
& \cdots .  \label{App_aDG_Convl}
\end{align}%
There are $C_{2M_{0}+1}^{i-1}$ sub-terms inside the summation in the $i^{\text{th}}$ term of Eq. (\ref{App_aDG_Convl}), $i=1,2,3,\cdots ,2M_{0}+1$. The second term is the summation of all the first order interactions. In the regime of $\mu \beta d T_0 \lesssim 1$ we can truncate Eq. (\ref{App_aDG_Convl}) at this level and neglect all the processes of re-absorption and re-emission among different sections of the medium.

Let us take $\omega _{m}=m\delta \omega $, where $\delta \omega >0$ is the frequency spacing between neighboring comb teeth. The summation in  the $2^\text{nd}$ term of Eq. (\ref{App_aDG_Convl}) can be written as
\begin{align}
\sum_{m=-M_{0}}^{M_{0}}e^{im\delta \omega \tau }& \approx \sum_{m=-\infty
}^{\infty }e^{im\delta \omega \tau }  \notag \\
& =T_{0}\sum_{n=-\infty }^{\infty }\delta \left( \tau -nT_{0}\right) ,
\label{App_exp_sum_delta}
\end{align}
where $T_{0}=2\pi /\delta \omega $. This is true if the duration of the GFC tooth in time domain $T_{0}/M$ is much smaller than the duration of the input field as well as the $e^{-\gamma t}e^{-i\beta t d/2}\,_{1}F_{1}(i\mu +1;2;i\beta dt)$ kernel. Namely, Eq. (\ref{App_exp_sum_delta}) requires a wide GFC bandwidth. Applying these approximations, we obtain the D-GFC echoes:
\begin{align}
a_{\text{DG}}^{(M_{0})} (t) \approx & \left(1-\frac{1}{2}|g|^{2}NdT_{0}\right)a_{\text{in}}(t) -\notag \\
& |g|^{2}NdT_{0} \sum_{n=1}^{\infty }a_{\text{in}}\left( t-nT_{0}\right) e^{-\gamma nT_{0}} \times  \notag \\
& \quad e^{-i\beta \frac{d}{2}nT_{0}}\Hypergeometric{1}{1}{i\mu
+1}{2}{i\beta d nT_0}.  \label{App_aDG2}
\end{align}

\subsection{Stepwise gradient frequency comb}
\label{AppSec_GFC1_sub2}
Similarly, from Eq. (\ref{App_as1_beta=0}), we have
\begin{equation}
a_{\text{SG}}^{(m)}(t)=\int_{-\infty }^{\infty }d\tau a_{\text{SG}
}^{(m-1)}(t-\tau )f_{\text{SG}}^{(m)}(\tau ),  \label{App_aSG1}
\end{equation}
where the subscript ``SG" denotes ``stepwise gradient", and
\begin{equation}
f_{\text{SG}}^{(m)}(t)=\delta (t)-|g|^{2}Nde^{i\omega _{m}t}e^{-\gamma t}
\frac{J_{1}(2\sqrt{|g|^{2}Ndt})}{\sqrt{|g|^{2}Ndt}}\Theta (t).
\end{equation}
From the recursion of Eq. (\ref{App_aSG1}), we obtain
\begin{align}
& a_{\text{SG}}^{(M_{0})}(t)  \notag \\
=& \int_{-\infty }^{\infty }d\tau _{0}\cdots d\tau _{2M_{0}}a_{\text{in}
}(t-\tau _{0}-\cdots -\tau _{2M_{0}})\times  \notag \\
& \qquad \qquad \qquad f_{\text{SG}}^{(-M_{0})}(\tau _{0})\cdots f_{\text{SG}
}^{(M_{0})}(\tau _{2M_{0}})  \notag \\
=& a_{\text{in}}(t)-|g|^{2}Nd\int_{-\infty }^{\infty }d\tau a_{\text{in}
}(t-\tau )e^{-\gamma \tau }\times  \notag \\
& \frac{J_{1}(2\sqrt{|g|^{2}Nd\tau })}{\sqrt{|g|^{2}Nd\tau }}\Theta (\tau
)\sum_{m=-M_{0}}^{M_{0}}e^{i\omega _{m}\tau }+\cdots .  \label{App_aSG_Convl}
\end{align}

Assuming that the duration of each comb tooth in time domain is smaller than the duration of the input field, $J_{1}(2\sqrt{|g|^{2}Ndt})/\sqrt{|g|^{2}Ndt}$, as well as the coherence lifetime: $\frac{T_0}{M} \lesssim \left\{ \Delta t,\frac{1}{|g|^{2}Nd},\frac{1}{\gamma }\right\} $, then we are able to apply the approximation (\ref{App_exp_sum_delta}), so that the S-GFC echoes read
\begin{align}
 a&_{\text{SG}}^{(M_{0})}(t)\approx  \left( 1- \frac{1}{2}|g|^{2}NdT_{0} \right) a_{\text{in}}(t)- |g|^{2}NdT_{0} \times \notag \\
&\sum_{n=1}^{\infty }a_{\text{in}}\left( t-nT_{0}\right)
e^{-\gamma nT_{0}}\frac{J_{1}(2\sqrt{|g|^{2}NdnT_{0}})}{\sqrt{
|g|^{2}NdnT_{0}}}.  \label{App_aSG2}
\end{align}
Eq. (\ref{App_aSG2}) is valid in the low optical thickness $|g|^{2}NdT_{0} \lesssim 1$ and wide comb bandwidth $\Delta
t>T_{0}/M $ regime.

\section{Analytical analysis of GFC for the first few echoes}
\label{AppSec_GFC2}
In this section, we use the Fourier transformation method on $t$ to calculate the first few D- and S-GFC echoes valid for general values of $|g|^2NdT_0$. The evolution equation of GFC are given in Sec. \ref{GFC} by Eqs. (\ref{EQDGa1}) and (\ref{EQDGS1}), where we consider $l_{m}=ml_{0}$, and define the ratio between $l_{0}$ and $d$ as $\mathcal{F}^{\prime }=l_{0}/d$, so that
\begin{equation}
L=d(M-1)\mathcal{F}^{\prime }+d.
\end{equation}%
For S-GFC, we take $\beta =0$, so the frequency spacing between the adjacent
comb tooth is $\delta \omega =\delta \omega ^{\prime }$.

In Fourier domain of $t$, Eqs. (\ref{EQDGa1}) and (\ref{EQDGS1}) are:
\begin{align}
\frac{\partial }{\partial z}a(z,\omega ) =&g^{\ast
}N\sum_{m=-M_{0}}^{M_{0}}S^{(m)}(z,\omega )\Theta ^{(m)}(z),
\label{App_EQDGaF1}
\end{align}
\begin{align}
-i\omega S^{(m)}(z,\omega ) =&-\left[ \gamma -i\Delta ^{(m)}(z)\right]
S^{(m)}(z,\omega )-  \notag \\
& ga(z,\omega ),  \label{App_EQDGSF1}
\end{align}
where the Fourier transformation is defined as
\begin{equation}
F(\omega )=\frac{1}{\sqrt{2\pi }}\int_{-\infty }^{\infty }F(t)e^{i\omega
t}dt.
\end{equation}
From Eq. (\ref{App_EQDGSF1}), we have
\begin{equation}
S^{(m)}(z,\omega )=-\frac{ga(z,\omega )}{\gamma -i\omega -i\Delta ^{(m)}(z)}.
\label{App_SDGF1}
\end{equation}
Substituting Eq. (\ref{App_SDGF1}) into Eq. (\ref{App_EQDGaF1}) and solving
for $a(z,\omega)$, we have
\begin{align}
& a(z,\omega ) =a_{\text{in}}(\omega ) \times  \notag \\
& \exp \left[ -\sum_{m=-M_{0}}^{M_{0}} \int_{-L/2}^{z} \frac{|g|^{2}N\Theta
^{(m)}(z^{\prime })}{\gamma -i\omega -i\Delta ^{(m)}(z^{\prime })}dz^{\prime
}\right] .  \label{App_aoutGFC_omega}
\end{align}

\subsection{Discontinuous gradient frequency comb}
\label{AppSec_GFC2_sub1}
From Eq. (\ref{App_aoutGFC_omega}), the D-GFC output field in Fourier domain is
\begin{align}
& a_{\text{DG}}(\omega )=a_{\text{in}}(\omega )\times  \notag \\
& \exp \left[ -|g|^{2}N\sum_{m=-M_{0}}^{M_{0}}\int_{-L/2}^{L/2}\frac{\Theta
^{(m)}(z^{\prime })}{\gamma -i\omega -i\Delta ^{(m)}(z^{\prime })}dz^{\prime
}\right] .  \label{App_aoutDGF1}
\end{align}
Now let us consider the function
\begin{equation}
F(\omega )=\sum_{m=-M_{0}}^{M_{0}}\int_{-L/2}^{L/2}\frac{\Theta
^{(m)}(z^{\prime })}{\gamma -i\omega -i\Delta ^{(m)}(z^{\prime })}dz^{\prime
}.
\end{equation}
Recalling that $\Delta ^{(m)}(z)=(m\delta \omega ^{\prime }+\beta z)\Theta^{(m)}(z)$, we have
\begin{align}
F(\omega )& =\sum_{m=-M_{0}}^{M_{0}}\int_{l_{m}-d/2}^{l_{m}+d/2}\frac{1}{
\gamma -i\omega -i(m\delta \omega ^{\prime }+\beta z^{\prime })}dz^{\prime }
\notag \\
& =\frac{i}{\beta }\sum_{m=-M_{0}}^{M_{0}}\ln \frac{\omega +\beta
d/2+m\delta \omega ^{\prime }+\beta l_{m}+i\gamma }{\omega -\beta
d/2+m\delta \omega ^{\prime }+\beta l_{m}+i\gamma },  \label{App_F_omega1}
\end{align}
where we consider $l_{m}=ml_{0}$. For the sake of simplicity, let us assume $\gamma \ll \beta d$. The $\delta \omega ^{\prime }$ term in Eq. (\ref{App_F_omega1}), if any, can be incorporated into $l_{0}$. So without loss of generality, it can be simply taken to be zero. Then the frequency spacing between nearby comb teeth is $\delta \omega =\beta l_{0}$, and $F(\omega)$ is written into
\begin{equation}
F(\omega )\approx \frac{i}{\beta }\sum_{m=-M_{0}}^{M_{0}}\ln \frac{\omega
+\beta d/2+m\beta l_{0}}{\omega -\beta d/2+m\beta l_{0}}.
\end{equation}
We assume that the medium opens a storage bandwidth well covering the input bandwidth of the signal, so that we are able to extend the summation limits from $\pm M_{0}$ to $\pm \infty $. In the regime $\mathcal{F}^{\prime } \gg 1$, $F(\omega )$ approximately becomes a periodic
function of $\omega$, which is equal to $\ln\frac{\omega +\beta d/2}{\omega -\beta d/2}$ in a period $\omega \in [-\beta l_0 /2, \beta l_0 /2]$. So we can write
\begin{align}
F(\omega ) & \approx \frac{i}{\beta }\left[ F_{0}+\sum_{n=1}^{\infty }A_{n}\cos \frac{2\pi
n\omega }{\beta l_{0}}+B_{n}\sin \frac{2\pi n\omega }{\beta l_{0}}\right] , \label{App_Fomega2}
\end{align}
where
\begin{equation}
F_{0}=\frac{1}{\beta l_{0}}\int_{-\beta l_{0}/2}^{\beta l_{0}/2}d\omega \ln
\frac{\omega +\beta d/2}{\omega -\beta d/2}=-i\pi \frac{d}{l_{0}},
\label{App_F0}
\end{equation}
\begin{align}
A_{n}=& \frac{2}{\beta l_{0}}\int_{-\beta l_{0}/2}^{\beta l_{0}/2}d\omega
\ln \frac{\omega +\beta d/2}{\omega -\beta d/2}\cos \frac{2\pi n\omega }{
\beta l_{0}}  \notag \\
=& -\frac{2i}{n}\sin \left( n\pi \frac{d}{l_{0}}\right) ,  \label{App_An}
\end{align}
\begin{align}
B_{n}=& \frac{2}{\beta l_{0}}\int_{-\beta l_{0}/2}^{\beta l_{0}/2}d\omega
\ln \frac{\omega +\beta d/2}{\omega -\beta d/2}\sin \frac{2\pi n\omega }{
\beta l_{0}}  \notag \\
\approx & \frac{2}{\beta l_{0}}\int_{-\infty }^{\infty }d\omega \ln \frac{
\omega +\beta d/2}{\omega -\beta d/2}\sin \frac{2\pi n\omega }{\beta l_{0}}
\notag \\
=& \frac{2}{n}\sin \left( n\pi \frac{d}{l_{0}}\right) .  \label{App_Bn}
\end{align}
Notice that in Eqs. (\ref{App_F0}) and (\ref{App_An}), the integrals depend on the branching of the function $\ln (z)$. Specifically, the results obtained by replacing $i\pi \rightarrow i\pi +2ik\pi $, $k=0,\pm 1,\pm 2,\cdots $ all could be possible solutions, while the correct one can be chosen based on their physical consequences on $a_{\text{DG}}$.
For example, simple criterions $|a_{\text{DG}}(\omega)|\leqslant |a_{\text{in}}(\omega )|$ and $a_{\text{DG}}(t<t_{\text{in}})=0$ ($t_{\text{in}}$ is the arrival time of the input signal $a_{\text{in}}$) have to be fulfilled by selecting the proper branch.

Substituting Eqs. (\ref{App_Fomega2})-(\ref{App_Bn}), Eq. (\ref{App_aoutDGF1}) becomes
\begin{align}
& a_{\text{DG}}(\omega )=a_{\text{in}}(\omega )\exp \left[ -|g|^{2}NF(\omega
)\right]   \notag \\
=& a_{\text{in}}(\omega )\exp \left[ -\mu \left( \frac{\pi }{\mathcal{F}
^{\prime }}+\sum_{n=1}^{\infty }\frac{2}{n}\sin \frac{n\pi }{\mathcal{F}
^{\prime }}e^{i\omega nT_{0}}\right) \right]   \notag \\
=& a_{\text{in}}(\omega )e^{-\frac{\pi \mu }{\mathcal{F}^{\prime }}
}\prod\limits_{n=1}^{\infty }\sum_{q=0}^{\infty }\left( -\frac{2\mu }{n}
\sin \frac{n\pi }{\mathcal{F}^{\prime }}\right) ^{q}\frac{e^{i\omega nqT_{0}}
}{q!},  \label{App_aDG_freq1}
\end{align}%
where $\mu =|g|^{2}N/\beta $, $\mathcal{F}^{\prime }=l_{0}/d$, $T_{0}=2\pi/(\beta l_{0})$. Expanding the summation of $q$ in Eq. (\ref{App_aDG_freq1}), we obtain:
\begin{align}
& a_{\text{DG}}(\omega )=a_{\text{in}}(\omega )e^{-\frac{\pi \mu }{\mathcal{F
}^{\prime }}}\prod\limits_{n=1}^{\infty }\bigg(1-\frac{2\mu }{n}\sin \frac{
n\pi }{\mathcal{F}^{\prime }}e^{inT_{0}\omega }+  \notag \\
& \,\frac{2\mu ^{2}}{n^{2}}\sin ^{2}\frac{n\pi }{\mathcal{F}^{\prime }}
e^{in2T_{0}\omega }-\frac{4\mu ^{3}}{3n^{3}}\sin ^{3}\frac{n\pi }{\mathcal{F}
^{\prime }}e^{in3T_{0}\omega }+  \notag \\
& \,\frac{2\mu ^{4}}{3n^{4}}\sin ^{4}\frac{n\pi }{\mathcal{F}^{\prime }}
e^{in4T_{0}\omega }-\frac{4\mu ^{5}}{15n^{5}}\sin ^{5}\frac{n\pi }{\mathcal{F
}^{\prime }}e^{in5T_{0}\omega }+\cdots \bigg).
\end{align}
It then becomes straightforward to calculate the relatively low-sequence echoes. For example, the leakage and the first five D-GFC echoes in time
domain are as follows:
\begin{widetext}
\begin{align}
a_{\text{DG}}(t)& =\exp \left( -\frac{\pi \mu }{\mathcal{F}^{\prime }}%
\right) \Bigg[a_{\text{in}}(t)-2\mu \sin \frac{\pi }{\mathcal{F}^{\prime }}%
a_{\text{in}}(t-T_{0})+\left( -\mu \sin \frac{2\pi }{\mathcal{F}^{\prime }}%
+2\mu ^{2}\sin ^{2}\frac{\pi }{\mathcal{F}^{\prime }}\right) a_{\text{in}%
}(t-2T_{0})+ \notag  \\
& +\left( -\frac{2\mu }{3}\sin \frac{3\pi }{\mathcal{F}^{\prime }}+2\mu
^{2}\sin \frac{\pi }{\mathcal{F}^{\prime }}\sin \frac{2\pi }{\mathcal{F}%
^{\prime }}-\frac{4\mu ^{3}}{3}\sin ^{3}\frac{\pi }{\mathcal{F}^{\prime }}%
\right) a_{\text{in}}(t-3T_{0})+ \notag \\
& +\left( -\frac{\mu }{2}\sin \frac{4\pi }{\mathcal{F}^{\prime }}+\frac{\mu
^{2}}{2}\sin ^{2}\frac{2\pi }{\mathcal{F}^{\prime }}+\frac{4\mu ^{2}}{3}\sin
\frac{\pi }{\mathcal{F}^{\prime }}\sin \frac{3\pi }{\mathcal{F}^{\prime }}%
-2\mu ^{3}\sin ^{2}\frac{\pi }{\mathcal{F}^{\prime }}\sin \frac{2\pi }{%
\mathcal{F}^{\prime }}+\frac{2\mu ^{4}}{3}\sin ^{4}\frac{\pi }{\mathcal{F}%
^{\prime }}\right) a_{\text{in}}(t-4T_{0})+ \notag \\
& +\bigg(-\frac{2\mu }{5}\sin \frac{5\pi }{\mathcal{F}^{\prime }}+\mu
^{2}\sin \frac{\pi }{\mathcal{F}^{\prime }}\sin \frac{4\pi }{\mathcal{F}%
^{\prime }}+\frac{2\mu ^{2}}{3}\sin \frac{2\pi }{\mathcal{F}^{\prime }}\sin
\frac{3\pi }{\mathcal{F}^{\prime }}-\mu ^{3}\sin \frac{\pi }{\mathcal{F}%
^{\prime }}\sin ^{2}\frac{2\pi }{\mathcal{F}^{\prime }}- \notag \\
& \qquad -\frac{4\mu ^{3}}{3}\sin ^{2}\frac{\pi }{\mathcal{F}^{\prime }}\sin
\frac{3\pi }{\mathcal{F}^{\prime }}+\frac{4\mu ^{4}}{3}\sin ^{3}\frac{\pi }{%
\mathcal{F}^{\prime }}\sin \frac{2\pi }{\mathcal{F}^{\prime }}-\frac{4\mu
^{5}}{15}\sin ^{5}\frac{\pi }{\mathcal{F}^{\prime }}\bigg)a_{\text{in}%
}(t-5T_{0})+\cdots \Bigg]. \label{App_DG_firstFive}
\end{align}
\end{widetext}
From Eq. (\ref{App_DG_firstFive}), the first D-GFC echo reads
\begin{equation}
a_{\text{DG}}^{1}(t)=-2\mu \sin \frac{\pi }{\mathcal{F}^{\prime }}e^{-\frac{
\pi \mu }{\mathcal{F}^{\prime }}}a_{\text{in}}(t-T_{0}).
\label{App_DG_firstEcho}
\end{equation}

\subsection{Stepwise gradient frequency comb}
\label{AppSec_GFC2_sub2}
The result of S-GFC was derived in Ref.~\cite{Zhang15TBPgamma}, which is briefly summarized in the following. From Eq. (\ref{App_aoutGFC_omega}), the
S-GFC output field in Fourier domain is
\begin{align}
& a_{\text{SG}}(\omega )=a_{\text{in}}(\omega )\times  \notag \\
& \exp \left[ -|g|^{2}N\sum_{m=-M_{0}}^{M_{0}}\int_{-L/2}^{L/2}\frac{\Theta
^{(m)}(z^{\prime })}{\gamma -i\omega -i\Delta ^{(m)}(z^{\prime })}dz^{\prime
}\right] ,
\end{align}
where $\Delta ^{(m)}(z)=m\delta \omega ^{\prime }\Theta ^{(m)}(z)=m\delta \omega \Theta ^{(m)}(z)$. Now let us consider the function
\begin{align}
H(\omega )& =\sum_{m=-M_{0}}^{M_{0}}\int_{-L/2}^{L/2}\frac{\Theta
^{(m)}(z^{\prime })}{\gamma -i\omega -im\delta \omega \Theta
^{(m)}(z^{\prime })}dz^{\prime }  \notag \\
& =\sum_{m=-M_{0}}^{M_{0}}\frac{d}{\gamma -i\omega -im\delta \omega }.
\end{align}
In the limit of high finesse $\mathcal{F}=\delta \omega /2\gamma \gg 1$ and wide bandwidth comb, $H(\omega )$ can be approximated as a periodic function~\cite{Zhang15TBPgamma}:
\begin{equation}
H(\omega )\approx id\left[ H_{0}+\sum_{n=1}^{\infty }C_{n}\cos \frac{2\pi
n\omega }{\delta \omega }+D_{n}\sin \frac{2\pi n\omega }{\delta \omega }
\right] ,
\end{equation}
where
\begin{align}
H_{0}& \approx -i\frac{\pi }{\delta \omega }, \\
C_{n}& \approx -i\frac{2\pi }{\delta \omega }e^{-\left\vert \frac{2\pi n}{
\delta \omega }\right\vert \gamma }, \\
D_{n}& \approx \frac{2\pi }{\delta \omega }\text{sign}\left( \frac{2\pi n}{
\delta \omega }\right) e^{-\left\vert \frac{2\pi n}{\delta \omega }
\right\vert \gamma }.
\end{align}

Similar as before, we then have the S-GFC echoes in Fourier domain:
\begin{align}
&a_{\text{SG}}(\omega ) =a_{\text{in}}(\omega )\exp \left[
-|g|^{2}N H(\omega )\right]  \notag \\
=&a_{\text{in}}(\omega )e^{-|g|^{2}Nd\frac{T_{0}}{2}} \times  \notag \\
& \prod\limits_{n=1}^{\infty }\sum_{q=0}^{\infty }\left(
-|g|^{2}NdT_{0}e^{-\gamma nT_{0}}\right) ^{q}\frac{e^{i\omega nqT_{0}}}{q!}.
\label{App_SG_frequency1}
\end{align}

From the inverse Fourier transformation of Eq. (\ref{App_SG_frequency1}) we
obtain the leakage and the first five S-GFC echoes in the following:
\begin{widetext}
\begin{align}
a_{\text{SG}}(t)& =\exp \left( -\frac{\pi }{4}\zeta _{\text{eff}}^{0}\right)
 \bigg\{a_{\text{in}}(t)-\frac{\pi }{2}\zeta _{\text{eff}}^{0}e^{-%
\frac{\pi }{\mathcal{F}}}a_{\text{in}}(t-T_{0})+\left[ -\frac{\pi }{2}\zeta
_{\text{eff}}^{0}+\frac{1}{2}\left( \frac{\pi }{2}\zeta _{\text{eff}%
}^{0}\right) ^{2}\right] e^{-\frac{2\pi }{\mathcal{F}}}a_{\text{in}%
}(t-2T_{0}) \notag  \\
& +\left[ -\frac{\pi }{2}\zeta _{\text{eff}}^{0}+\left( \frac{\pi }{2}\zeta
_{\text{eff}}^{0}\right) ^{2}-\frac{1}{6}\left( \frac{\pi }{2}\zeta _{\text{%
eff}}^{0}\right) ^{3}\right] e^{-\frac{3\pi }{\mathcal{F}}}a_{\text{in}%
}(t-3T_{0})+ \notag \\
& +\left[ -\frac{\pi }{2}\zeta _{\text{eff}}^{0}+\frac{3}{2}\left( \frac{\pi
}{2}\zeta _{\text{eff}}^{0}\right) ^{2}-\frac{1}{2}\left( \frac{\pi }{2}%
\zeta _{\text{eff}}^{0}\right) ^{3}+\frac{1}{24}\left( \frac{\pi }{2}\zeta _{%
\text{eff}}^{0}\right) ^{4}\right] e^{-\frac{4\pi }{\mathcal{F}}}a_{\text{in}%
}(t-4T_{0})+ \notag \\
& +\left[ -\frac{\pi }{2}\zeta _{\text{eff}}^{0}+2\left( \frac{\pi }{2}\zeta
_{\text{eff}}^{0}\right) ^{2}-\left( \frac{\pi }{2}\zeta _{\text{eff}%
}^{0}\right) ^{3}+\frac{1}{6}\left( \frac{\pi }{2}\zeta _{\text{eff}%
}^{0}\right) ^{4}-\frac{1}{120}\left( \frac{\pi }{2}\zeta _{\text{eff}%
}^{0}\right) ^{5}\right] e^{-\frac{5\pi }{\mathcal{F}}}a_{\text{in}%
}(t-5T_{0})+\cdots \bigg\}, \label{App_SG_firstFive}
\end{align}
\end{widetext}
where we introduce the individual effective optical thickness $\zeta _{\text{eff}}^{0}=2|g|^{2}NdT_{0}/\pi $.

The first S-GFC echo takes the form of
\begin{equation}
a_{\text{SG}}^{1}(t)=-\frac{\pi \zeta _{\text{eff}}^{0}}{2}e^{-\frac{\pi
\zeta _{\text{eff}}^{0}}{4}}e^{-\frac{\pi }{\mathcal{F}}}a_{\text{in}%
}(t-T_{0}),  \label{App_SG_firstEcho}
\end{equation}
with corresponding efficiency given by Eq. (\ref{eff_SG1}).
When the optimization condition
\begin{equation}
\zeta _{\text{eff}}^{0}=\frac{4}{\pi }
\end{equation}
is satisfied, the efficiency reaches its maximum value $\approx 54\%$ if $\mathcal{F}\gg 1$.

By introducing the effective gradient $\tilde{\beta}$ and spacing $l_{0}$ in the S-GFC regime, we are able to define $\mu =|g|^{2}N/\tilde{\beta}$ and $\mathcal{F}^{\prime }=\tilde{l}_{0}/d$ in the same way as in the D-GFC regime. The individual effective optical thickness then becomes $\zeta _{\text{eff}}^{0}=4\mu /\mathcal{F}^{\prime }$. Therefore we can write the first S-GFC echo (\ref{App_SG_firstEcho}) into the following form
\begin{equation}
a_{\text{SG}}^{1}(t)=-2\mu \frac{\pi }{\mathcal{F}^{\prime }}e^{-\frac{\pi
\mu }{\mathcal{F}^{\prime }}}e^{-\frac{\pi }{\mathcal{F}}}a_{\text{in}
}(t-T_{0}),  \label{App_SG_firstEcho2}
\end{equation}
which makes a good comparison with the first D-GFC echo given by Eq. (\ref{App_DG_firstEcho}).

\bibliographystyle{apsrev4-1}
\bibliography{QMGEMANAbibFile}

\end{document}